\documentclass[twocolumn,aps,prb,floats]{revtex4} 
\usepackage{amssymb,amsbsy,amsmath,mathrsfs}
\usepackage{epsfig,psfrag,subfigure}
\usepackage{graphicx,float}
\usepackage{times}
\usepackage{color}
\usepackage{subfigure}
\usepackage{setspace}
\usepackage{bm} 
\usepackage{xr}
\usepackage{stackengine}
\usepackage{natbib,url,hyperref}
\hypersetup{colorlinks,allcolors=blue}
\usepackage{filecontents}
\usepackage{comment}

\newcommand\bea{\begin{eqnarray}}
\newcommand\eea{\end{eqnarray}}
\newcommand\beq{\begin{equation}}
\newcommand\eeq{\end{equation}}

\newcommand{\noi}{\noindent}
\newcommand{\non}{\nonumber}
\newcommand{\al}{\alpha}

\newcommand{\ga}{\gamma}

\newcommand{\ra}{\rangle}
\newcommand{\uaw}{\uparrow}

\def\red{\textcolor{red}}

\newcommand{\ket}[1]{|#1\rangle}
\begin{document}
\title{
Subspace-restricted thermalization in a correlated-hopping model with strong Hilbert 
space fragmentation characterized by irreducible strings}
\author{Sreemayee Aditya$^1$, Deepak Dhar$^{2} \footnote{This author contributed to
Sections~I-III and Appendices A-C.}$, and Diptiman Sen$^1$}
\affiliation{$^1$Centre for High Energy Physics, Indian Institute of Science,
Bengaluru 560012, India \\
$^2$ Indian Institute of Science Education and Research,
Dr. Homi Bhabha Road, Pashan, Pune 411008, India}

\begin{abstract}

We introduce a one-dimensional correlated-hopping model of spinless fermions 
in which a particle can hop between two neighboring sites only if the sites to 
the left and right of those two sites have different particle numbers.
Using a bond-to-site mapping, this model involving four-site terms can be 
mapped to an assisted pair-flipping model involving only three-site terms. 
This model shows strong Hilbert space fragmentation (HSF). We define irreducible 
strings (IS) to label the different fragments, determine the number of fragments, 
and the sizes of fragments corresponding to some special IS. 
In some classes of fragments, the Hamiltonian can be diagonalized completely, and in others it can be seen to have a structure characteristic of models
which are not fully integrable.
In the largest fragment in our model, the number of states grows exponentially
with the system size, but the ratio of this number to the total Hilbert space
size tends to zero exponentially in the thermodynamic limit. 
Within this fragment, we provide numerical evidence that only a weak version of 
the eigenstate thermalization hypothesis (ETH) remains valid; we call this 
subspace-restricted ETH.
To understand the 
out-of-equilibrium dynamics of the model, we study the infinite-temperature 
time-dependent autocorrelation functions starting from a random initial state; 
we find that these exhibit a different behavior near the boundary compared to the bulk.
Finally we propose an experimental setup to realize our correlated-hopping model.
\end{abstract}

\maketitle

\section{Introduction}

Thermalization and its violation in isolated quantum systems have
been studied extensively over the last several years. The strongest version of 
thermalization in closed quantum systems is believed to be defined by the eigenstate thermalization hypothesis (ETH)~\cite{sredniki_1994,deutsch_1991,rigol_2008,polkovnikov_2011}. This 
hypothesis states that each eigenstate in an ergodic system acts like a thermal ensemble as far as local observables are concerned, namely, local correlation 
functions in each eigenstate tend to the average values calculated in the quantum mechanical microcanonical
ensemble at the same energy. Thus expectation values of local observables for all eigenstates for a sufficiently {\em large} system show 
``self-thermalization" on their own.

This can be further categorized into two classes called the strong and weak versions 
of ETH. In the first case, all of the eigenstates of a given Hamiltonian satisfy the
ETH hypothesis, while in the latter case, it is obeyed by most of the eigenstates 
apart from some states which form a set of measure zero in the thermodynamic limit.
Well-known examples where strong ETH is not valid are quantum integrable models 
which are either mappable to free systems\cite{freefermi1,freefermi2} or are solvable by the Bethe ansatz~\cite{bethe1,bethe2,bethe3}. Strong ETH is also not 
satisfied in many-body localized systems which have strong disorder~\cite{pal_2010,vosk_2013,huse_2013,nandokishore_2015}.

In recent years, there have been a great deal of effort in identifying systems for which the strong version of ETH is not valid. One class of systems is
``quantum many-body scars"~\cite{iadecola_2020,turner_2018,choi_2019,
moudgalya_2018,lin_2019,iadecola_2019}. Such systems have some special states
called scar states which are highly excited and have
low entanglement-entropy. If an initial state, which has a 
significant overlap with such eigenstates compared to other states at the same energy, is evolved in
time, the long-time dynamics of correlation functions show persistent oscillations. For an quantum ergodic system, such overlaps should be a smooth function of 
energy; therefore, the enhancement of overlaps for certain eigenstates compared to others is not consistent with the strong ETH.
However, the presence of such states still satisfies the weak ETH with respect to the full Hilbert space since the number of scar states typically grows 
only polynomially with the system size.

More interesting are systems that show Hilbert space fragmentation (HSF)~\cite{Moudgalya_review_2022}, which we will focus on 
in this paper. In these systems, the Hamiltonian takes a block-diagonal form in a basis given by a product of local states, and the total number of blocks 
increases exponentially with the size of the system. This differs 
from what occurs when there are a finite number of global symmetries; if there is a finite number $r$ of conserved quantities in the system, the number of blocks 
grows polynomially as $V^r$, where $V$ is the volume of the system. 
We emphasize that if there is HSF, the ETH fails in general with respect to the total Hilbert space. Another striking feature is the 
presence of frozen states which are basis vectors that are eigenstates of the Hamiltonian with zero energy. The number of such frozen states grows 
exponentially with the system size. The most interesting kind of HSF is strong 
fragmentation, where the size of the largest fragment is exponentially smaller 
than the full many-body Hilbert space. 

There are some challenges in understanding different aspects of fragmented systems where the constants of motion cannot be expressed as integrals of local 
observables. Some important steps towards a complete characterization are the 
concepts of ``statistically localized integrals of 
motion"~\cite{sala_stat_loc_2020} (which is equivalent to our construction of IS),
commutant algebras~\cite{moudgalya_PRX_2022}, and the algebra of matrix product 
operators~\cite{pozsgay_2023}; these uniquely label the disconnected 
subspaces in several fragmented systems. Moreover, the construction of commutant algebras further categorizes this mechanism based on the basis states in 
which fragmentation takes place. If the fragmentation occurs in a 
particle number basis, it is called classical 
fragmentation~\cite{Moudgalya_pairhopping_2020,sala_ergo_2020}, 
whereas fragmentation happening in an entangled basis is dubbed as secondary
fragmentation~\cite{mukherjee_2021} or quantum 
fragmentation~\cite{brighi_2023,moudgalya_PRX_2022}. The model studied in our
paper will turn out to show classical fragmentation.

Examples of such systems for which the strong ETH is not valid are systems with dipole 
conservation or conserved magnetization~\cite{tomasi_2019,Moudgalya_review_2022,
Moudgalya_pairhopping_2020,khemani_2020,sala_ergo_2020,lenart_2021,pozsgay_2021,
pozsgay_2023}. In these systems, HSF occurs due to strong constraints on the 
mobility of excitations. However, there are also examples of HSF which do not involve dipole 
conservation~\cite{yang_2020,brighi_2023,lee_2021,mukherjee_2021,alexy_2022,atsuki_2022,Hart_2022,stephen_2024}. 
There are also some studies of periodically driven models where HSF has been 
found recently~\cite{ghosh_2023,aditya_2023,zhang2023floquet}. 

In this work, we introduce and study a one-dimensional correlated-hopping model of spinless fermions with terms informing four consecutive sites.
Using a bond-site mapping to a dual lattice~\cite{lenart_2021,pozsgay_2021}, 
this model can be mapped to an assisted pair-flipping model that only has 
terms involving three sites. We find that this system shows strong HSF.

To characterize the HSF in this system, we use the idea of ``irreducible strings" 
(IS). The concept of IS was introduced several years ago to understand 
exponential fragmentation (called ``many-sector decomposition" at that time)
in several classical models such as a deposition-evaporation model of 
$k$-mers~\cite{Barma_1994,HariMenon_1995} and a diffusing dimer model~\cite{menon_1997}.
Recently, the idea of IS has been rediscovered and called `statistically localized 
integrals of motion'. In particular, Rakovszky et al \cite{sala_stat_loc_2020} 
have used the idea of statistically localized integrals of motion to study 
various systems exhibiting strong HSF, 
and they have also shown that correlation functions have a non-uniform profile 
whose value near the boundary does not agree with the microcanonical 
expectation value.

Remarkably, many features of our model can
be understood in terms of IS, such as the total number of fragments and total s
number of frozen states. We compute the size of the largest fragment
employing the idea of IS and an enumerative combinatorics of characters~\cite{generating1994,HariMenon_1995,menon_1997,Barma_1994}, which we 
have
verified by direct numerical checks. We find that strong ETH is not satisfied in
our model, as suggested by an analysis of the energy level spacing 
ratio~\cite{Wigner_1955,serbyn_2016,pal_2010} and expectation values of few-body operators of all the eigenstates (without resolving into fragments), which is 
always 
the case for a typical fragmented system~\cite{sala_ergo_2020}. A similar
analysis for the largest fragment indicates that this subspace is 
non-integrable. We provide evidence that a weaker form of subspace-restricted 
ETH still holds within sufficiently large fragments which we call typical fragments~\cite{Moudgalya_pairhopping_2020,moudgalya_PRX_2022}.

Next, we study the out-of-equilibrium dynamics in our model to look for dynamical signatures of the lack of thermalization. We find that
infinite-temperature autocorrelation functions starting from a random initial state in the full Hilbert space also show that strong ETH is not
satisfied. Moreover, the boundary autocorrelation 
function oscillates around a finite saturation value at long times, which is much 
larger than the bulk saturation value~\cite{sala_ergo_2020,sala_stat_loc_2020,moudgalya_PRX_2022}. We provide an
understanding of the non-uniform profiles of the bulk and boundary spectra
by computing the lower bound of these two autocorrelation functions using the Mazur 
inequality~\cite{MAZUR1969533,SUZUKI1971277} and a knowledge of the fragmentation structure of the model. We also study the entanglement dynamics in the largest fragment. This confirms our previous 
finding that ETH is not satisfied in the full Hilbert space, but a weak
subspace-restricted ETH is valid within the largest fragment. 

We conclude by presenting an experimentally realizable
$t-V$ model with a spatial periodicity of four which can generate our
correlated-hopping model in a particular limit. Another way to realize
our model is through a periodically driven system with an on-site potential
with a spatial periodicity of four sites. We find that for some particular
driving parameters, an interplay between dynamical localization (i.e.,
the effective hopping becoming zero as a result of the driving), resonances 
between different states, and density-density interactions gives rise to 
precisely this model~\cite{aditya_2023}.

The plan of this paper is as follows. In Sec.~\ref{sec2}, we discuss the 
Hamiltonian of our model, its global symmetries, and a mapping to a model
with three-site terms. In Sec.~\ref{sec3}, we consider the three-site 
form of the model, and we discuss the fragmentation of
the Fock space, the number of fragments, the number
of frozen fragments which contain only one state each, and a description of
some special fragments, including the largest sector. Some details are
relegated to the Appendices. In Secs.~\ref{sec4}-\ref{sec7},
we consider the four-site form of the model. In Sec.~\ref{sec4}, 
we provide evidence for ergodicity breaking and non-integrability
through the expectation values of some local operators, the half-chain
entanglement entropy, and the distribution of the energy level spacing.
In Sec.~\ref{sec5}, we study dynamical signatures of HSF by looking at the
long-time behavior of autocorrelation functions and the time evolution of
the half-chain entanglement entropy. In Sec.~\ref{sec6}, we discuss how our
model may arise in the large-interaction limit of a variant of the $t-V$
model in which nearest-neighbor interactions have a period-four structure.
For the purposes of comparison, we discuss in Sec.~\ref{sec7} a different
model with four-site terms which has been studied extensively in recent years as 
an example of a system exhibiting HSF. We show that this can be mapped to a 
model with three-site terms which describes stochastic evolution of
diffusing dimers on a line. The complete structure of the HSF in the latter
model was found exactly many years ago using the idea of IS~\cite{menon_1997}.
We summarize our results and point out some directions for future studies 
in Sec.~\ref{sec8}.

\section{Model Hamiltonian and symmetries}
\label{sec2}

We consider a one-dimensional spinless fermionic model which, for an 
infinitely large system, is described by the Hamiltonian
\bea H_1 ~=~ \sum_{j} ~(n_{j}-n_{j+3})^{2}\,(c_{j+1}^{\dagger}c_{j+2}
+ c_{j+2}^{\dagger}c_{j+1}). \label{ham1} \eea
Here $c_{j} ~(c_j^\dagger)$ is a fermionic annihilation (creation) operator on site $j$, and $n_{j}=c_{j}^{\dagger}c_{j}$ can take values 0 or 1. This 
Hamiltonian connects the following pairs of states involving four consecutive sites,
\bea 1100 &\leftrightarrow& 1010 \non\\
0101&\leftrightarrow &0011. \label{proori} \eea
We define a spin variable $Z_{j}=2n_{j}-1$ which can only take the values 
$\pm 1$ at site $j$. Apart from translation and inversion symmetries, this
model has three additional global symmetries: total particle number $C_1$, 
and two staggered quantities $C_{2}$ and $C_{3}$ given by
\bea C_{1} &=& \sum_j ~n_{j}, \non\\
C_{2} &=& \sum_j ~(-1)^{j} ~Z_{2j} Z_{2j+1}, \non\\
C_{3} &=& \sum_j ~(-1)^{j}Z_{2j-1} Z_{2j}. \label{globalsym} \eea
Moreover, at half-filling, this model is invariant under a modified particle-hole 
transformation given by $c_{j}\leftrightarrow (-1)^j ~c_{j}^{\dagger}$.
It turns out that these are only three of many other conserved quantities which
can be characterized in terms of a construct called irreducible strings; this will
be discussed in Sec.~\ref{sec3}.

The energy spectrum of this model has an $E \to - E$ symmetry. This can be seen by noting that the 
transformation $c_j \to (-1)^j c_j$ and $c^\dagger_j \to (-1)^j c^\dagger_j$ changes the Hamiltonian $H \to - H$.

The model in Eq.~\eqref{ham1} can be mapped to a different model with a Hamiltonian in which the degrees of freedom lie on the bonds of the original 
lattice, and the new
Hamiltonian is a sum of terms involving only three consecutive 
sites; this makes it easier to study the system. 
We map states for a bond $(j,j+1)$ to a 
state on the site $j+1/2$ on the dual lattice following the rules,
\bea |01 \rangle &\to& | 1 \rangle, ~~~~|10 \rangle ~\to~ | 1 \rangle, \non \\
|00 \rangle &\to& | 0 \rangle, ~~~~|11 \rangle ~\to~ | 0 \rangle. \label{2to1} \eea
This is clearly a two-to-one mapping from the four-site model to the three-site
mode on the dual lattice.
For example, the two states $10101010$ and $01010101$ of 
the four-site model (these states are related to each other by a particle-hole
transformation) map to a single state $1111111$ of the three-site model.

\begin{figure}[h!]
\includegraphics[width=\columnwidth]{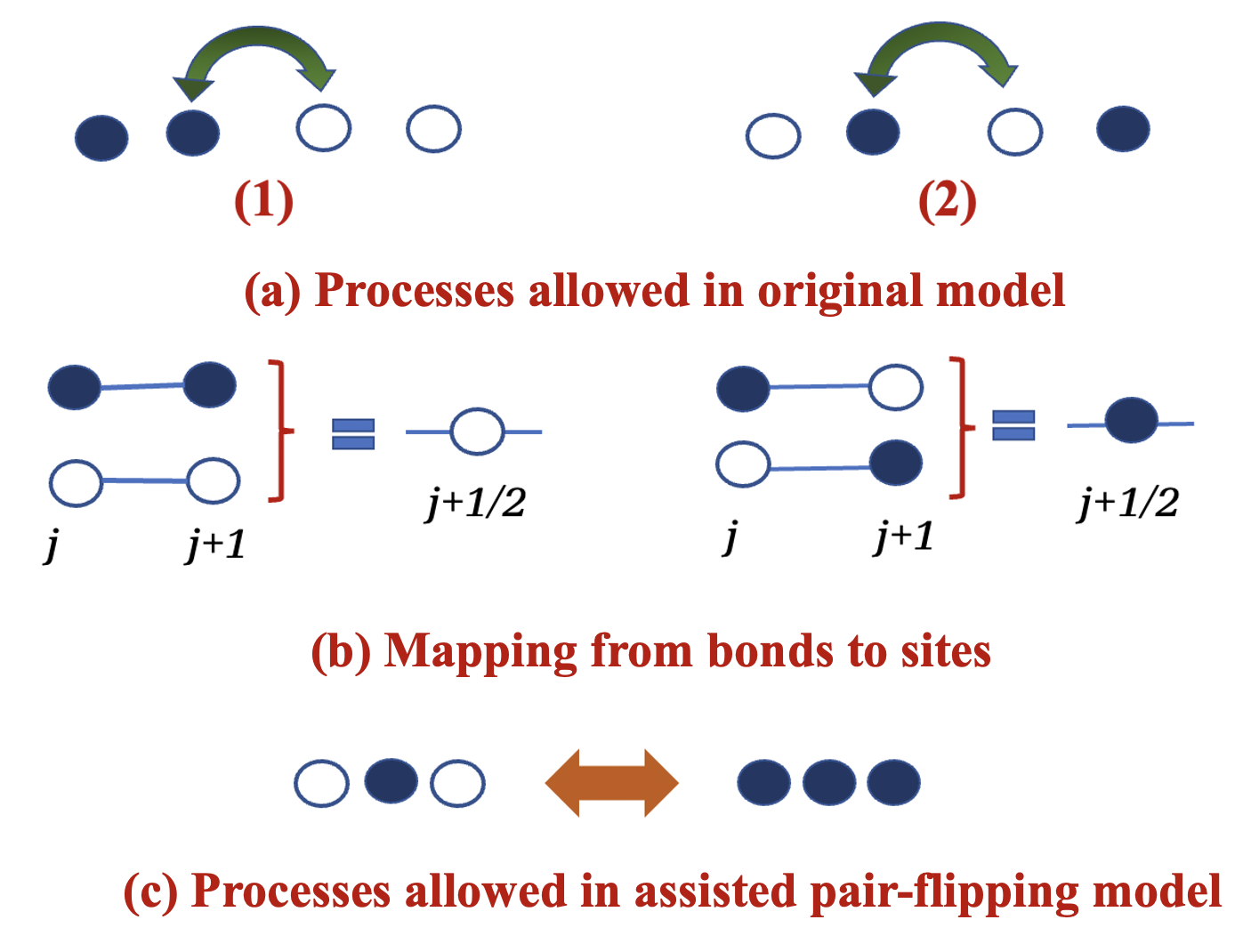}%
\caption{(a) Processes allowed in original correlated-hopping model with 
terms involving four consecutive sites. (b) Mapping of bonds to sites.
(c) The bond-site mapping converts the correlated-hopping model to an assisted 
pair-flipping model with terms involving three consecutive sites. Here 
filled and empty circles represent occupied and unoccupied sites respectively.}
\label{model} \end{figure}

We see that the three-site model has two states $0$ and $1$ 
at each site, and Eq.~\eqref{proori} implies that only the following transitions 
are allowed for this model,
\bea 111 \leftrightarrow 010. \label{trans1} \label{dualtra} \eea
The model is therefore described by a Hamiltonian which involves only three 
consecutive sites of the dual lattice,
\bea H_2 ~=~ \sum_j ~n_{j+1}~ (d_{j}^{\dagger}d^{\dagger}_{j+2} ~+~ d_{j+2} d_j),
\label{ham2} \eea
where $n_j = d_j^\dagger d_j$. This rule implies that a pair
of spinless fermions can be created or annihilated on two 
next-nearest-neighbor sites provided that the site in the middle is occupied. It is 
important to note that this three-site Hamiltonian
does not conserve the total particle number unlike the four-site model.
A summary of the original and final Hamiltonians in Eqs.~\eqref{ham1} and 
\eqref{ham2} and the bond-site mapping connecting the two is shown in Fig.~\ref{model}.

\section{Fragmentation of the Hilbert space}
\label{sec3}

In this section, we will show how kinetic constraints in our model shatters the 
Hilbert space leading to an exponentially large number of fragments in the local 
number basis. The fragmentation structure can appear in a variety of forms, such 
as frozen fragments 
consisting of a single eigenstate of the Hamiltonian and 
fragments with a finite or exponentially large number of states.

\subsection{Irreducible strings}
\label{irreducible}

We first discuss the fragmentation structure in the three-site model which 
allows the transition described in Eq.~\eqref{dualtra}; 
the analytical treatment is much simpler in the three-site language. It is 
convenient to define a Hamiltonian which only has matrix elements between the 
states $010$ and $111$. We therefore define Pauli matrices $\sigma_{j}^{\alpha}$ at site $j$ (where $\alpha=x,y,z$), so that states 1 and 0 at site j corresponds 
to $\sigma_{j}^{z}=\pm 1$. In terms of these Pauli spin operators, the
Hamiltonian becomes
\bea H_3 =\sum_{j}~(\sigma_{j-1}^{+}\sigma_{j+1}^{+}+\sigma_{j-1}^{-}
\sigma_{j+1}^{-}) \,\left(\frac{1+\sigma_{j}^{z}}{2}\right). \label{assispf} \eea
Note that this expression differs from $H_2$ in that it is defined in terms of Pauli spin operators which commute at different sites.
This is an assisted spin-flipping Hamiltonian in which a pair of spins can flip
on next-nearest-neighbor sites provided the site in the middle has spin
$\uaw$. The number of states in a system with $L$ sites is $2^{L}$.
We will consider open boundary conditions (OBC) to perform our analysis of fragmentation.
One should note that the transition rules shown in Eq.~\eqref{assispf} imply that 
\beq 11110 \leftrightarrow 01010 \leftrightarrow 01111. \label{move} \eeq
This implies that a string of four 1's can move 
across either a 0 or a 1 (trivially). This will be important later 
to understand different features of the HSF occurring in this model.

We will show below that the model in Eq.~\eqref{assispf} has an exponentially large number of fragments. These fragments are most easily characterized in terms of a 
construction called IS which act as an exponentially large number of
conserved quantities. This construction of IS is a variation of the construction used earlier in Refs.~\onlinecite{HariMenon_1995,menon_1997,Barma_1994}. This is defined as follows.

\begin{figure}[h!]
\includegraphics[width=\columnwidth]{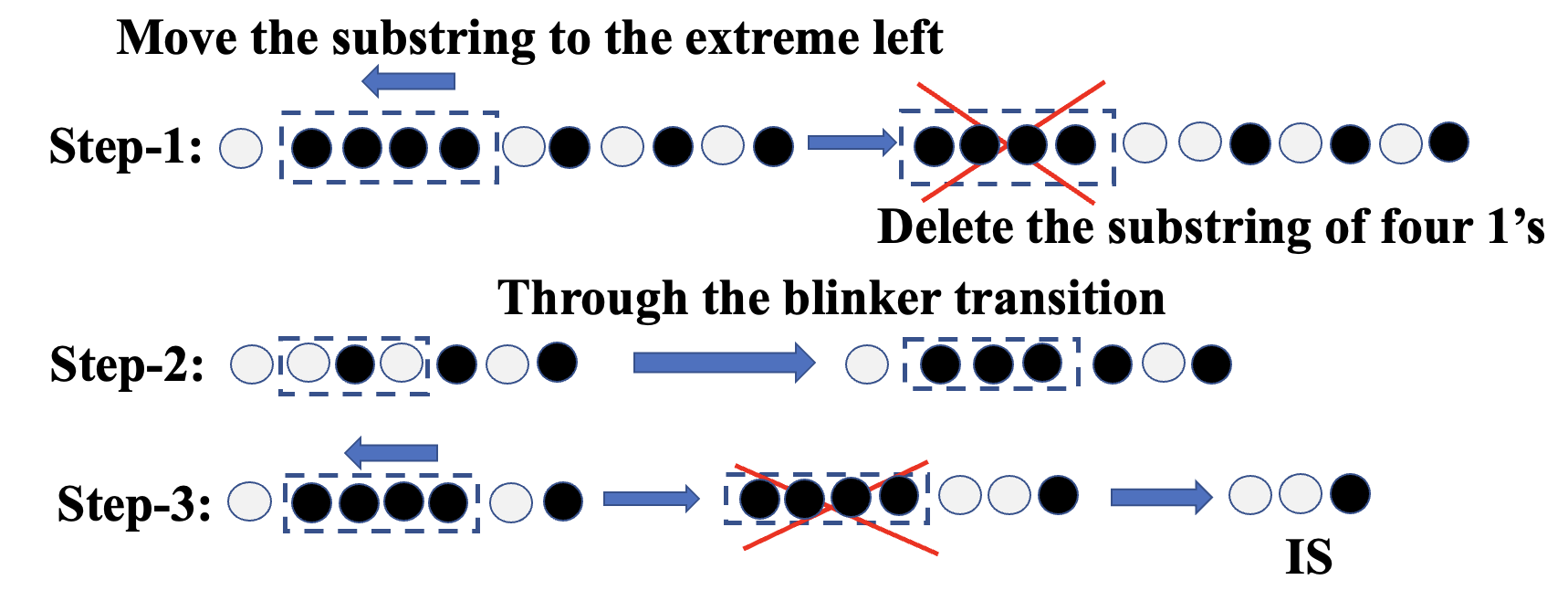}%
\caption{Illustration of IS construction for a given binary string configuration.}
\label{IS_example} \end{figure}

A basis configuration is characterized by a binary string of length $L$, e.g., 
$0011010 \cdots$.
We read the string from left to right, move the first occurrence of $1111$ to
the left end of the string and then delete it. This reduces the length of the 
string by $4$. We repeat this till no further reduction is possible. 
Then, we read the remaining string from left to right, and change the first occurrence of $010$ to $111$. If this generates a $1111$, we move it to the leftmost end and delete this to get a string of 
reduced length. The steps $1111 \rightarrow \phi$ (null string), and $010 \rightarrow 111$ are repeated, till no further change can be made. The final string is the IS 
corresponding to the initial string. As an example, one can see that the IS for the string configuration $01111010101$ reduces to $001$ by the rules mentioned 
above; this is depicted in Fig. \ref{IS_example}.

The usefulness of the IS construction comes from the observation that two states belong to the same fragment if 
and only if they have the same IS. To prove this assertion, we note that the strings $01111$ and $11110$ are obtainable from each other as shown in Eq.~\eqref{move}. Hence 
one may treat a group of four adjacent $1$'s as a block that can slide across a $0$, and it can slide across a $1$ 
trivially. Thus, we can push any such blocks of $1111$ to the left end of the
system and delete them.

An IS of length $L- 4r$ corresponds to a root state of length $L$ which has
$4r$ $1$'s at the left. We also note that for each of the steps going from the initial string to the root state, the inverse steps are also allowed through the transitions 
$010
\leftrightarrow 111$. Hence, if two configurations have the same root state, they can be reached from each other. The Hilbert space fragment corresponding to a given IS is
spanned by all the configurations that have that IS. Thus the IS acts as a unique label for the fragment.

\subsection{Determining the number of fragments}
\label{counting}

To calculate the number of fragments, it is convenient to define a number $M(n)$
which is the number of distinct IS of length $n$. We know that an IS cannot contain
the substrings $010$ or $1111$ anywhere. Using this fact, we can calculate $M(n)$,
for $n > 3$, using a transfer matrix method as shown in Appendix~\ref{appml}.

\begin{table}[h!]
\begin{center}
\begin{tabular}{|c|c|c|c|c|c|c|c|c|c|c|c|c|c|} 
\hline
$n$& 0& 1 &2 &3 &4 &5 &6 &7 &8 &9 &10&11&12\\
\hline
$M(n)$&1&2&4&7&11&18&29&47&76&123&199&322&521\\
\hline
\end{tabular}
\end{center}
\caption{Values of $M(n)$ versus $n$.}\label{Table I}
\end{table}

It is convenient to define $M(0)=1$. Clearly, $M(1)=2$, $M(2)=4$ and $M(3)=7$. 
The results for the first few values of $M$ are given in Table I.
Given $M(n)$, the number of fragments of length $L$ with OBC is given by
\bea N_{OBC}(L)~=~ \sum_{r =0}^{\lfloor L/4\rfloor} M(L-4 r), {\rm ~ for ~} L>4.\non\\
\label{nobc}\eea
The resulting values are used to analytically calculate the $N_{OBC}(L)$, given in Table II; we have verified these values numerically. We also list the corresponding 
values for PBC found numerically (using a method described in Appendix~\ref{appml}). We find that both $N_{OBC}$ and $N_{PBC}$ grow asymptotically as $\tau^L$, where 
$\tau = (\sqrt{5}+1)/2 \simeq 1.618$ is the golden ratio.

\begin{table}[ht] 
\begin{center}
\begin{tabular}{|c|c|c|c|c|c|c|c|c|c|} 
\hline
$L$ & 4 & 5 & 6 & 7 & 8 &9 & 10 & 11 & 12 \\
\hline
$N_{\rm OBC}$ & 12 & 20 & 33 & 54 & 88 & 143 & 232 & 376 & 609 \\
\hline
$N_{\rm PBC}$ & 10 & 13 & 20 & 32 & 59 & 81 & 131 & 207 & 363 \\
\hline
\end{tabular}
\end{center}
\caption{Values of $N_{OBC}$ and $N_{PBC}$ versus $L$ obtained by numerical enumeration.}
\label{Table II}
\end{table}

\subsection{Description of frozen fragments}
\label{sec3c}

The model contains an exponentially large number of eigenstates that do not participate in the 
dynamics; we call these ``frozen" states. The frozen states $\ket{f_{j}}$ are all 
product states in the particle number basis, and are annihilated by the Hamiltonian so that $H\ket{f_{j}}=0$. Hence these states are 
zero-energy eigenstates of the model Hamiltonian. Two trivial examples of such states in the four-site language are fully empty and fully occupied states in the particle number
basis, i.e., $\ket{\cdots 0000 \cdots}$ and $\ket{\cdots 1111 \cdots}$, respectively. 

In a frozen state, there cannot be any occurrence of substrings $010$ or $111$, and the length of the IS must be $L$. It is then straightforward to set up a transfer 
matrix to determine the exact numbers of such states $N^{frozen}(L)$ for a system of size $L$.
The details of the calculation are given in Appendix~\ref{appfrozen}. 
We find that the number of frozen fragments for a system of size $L$ grows as $1.466^L$ for large $L$.
For the first few values of $L$, the number of frozen states $N^{frozen}(L)$
is shown in Table III for both OBC and PBC.

\begin{table}[h!]
\begin{center}
\begin{tabular}{|c|c|c|c|c|c|c|c|c|c|c|} 
\hline
$L$&3 &4 &5 &6 &7 &8 &9 &10&11&12\\
\hline
$N^{frozen}_{\rm OBC}$&6&9&13&19&28&41&60&88&129&189\\
\hline
$N^{frozen}_{\rm PBC}$&4&5&6&10&15&21&31&46&67&98\\
\hline
\end{tabular}
\end{center}
\caption{$N^{frozen}_{\rm OBC}$ and $N^{frozen}_{\rm PBC}$ versus $L$ with OBC and PBC
respectively.} \label{Table III}
\end{table}

We note here that the frozen states possess many of the features 
of quantum many-body scars. They have exactly zero energy (and hence lie in the 
middle of the energy spectrum), and this does not change
if the coefficients of the terms in the Hamiltonian in Eq.~\eqref{ham1} are
taken to be random numbers instead of all being set equal to $1$. Further, 
since the frozen states are all product states in real space, they trivially
have zero entanglement entropy between any two parts of a system with either OBC or PBC.

\subsection{Description of some simple integrable fragments}
\label{simple}

We will now present some examples of fragments which consist of more
than one state (and are therefore not frozen) but in which the Hamiltonian
dynamics is integrable.

The first example of an integrable fragment is a set of multiple "blinkers" 
each of which flips back and forth between two states. For example, we can have an 
IS of the form
$$\cdots 00011100111000 \cdots$$ 
These fragments consist of a sea of $0$'s with islands of three consecutive sites that can flip between $010$ and $111$ but are fully 
localized in space. The general state in such a fragment can be obtained
by concatenating the substrings $0$, $1110$ and $0100$. For a system with
OBC, the number of states in the fragment can be found by defining a transfer
matrix $T$ following a procedure similar to the one given in Appendix~\ref{appml}.
We find that $T$ is a $8 \times 8$ matrix whose characteristic polynomial is
given by $\lambda^5 (\lambda^3 - \lambda^2 - 1)$. The largest root of this equation
is approximately $1.466$ (see Eqs.~\eqref{lambda1} and \eqref{lambda2}), and 
the number of states 
therefore grows with system size as $1.466^L$ asymptotically. For each blinker,
labeled by an integer $j=1,2,\cdots,r$, we can introduce a Pauli matrix $\tau_j^z$
which is equal to $\pm 1$ for $111$ and $010$ respectively. The number of states 
in this fragment is equal to $2^r$, and the effective Hamiltonian
is given by $H_{eff} = \sum_{j=1}^r \tau_j^x$. It is then easily seen that the energy eigenvalues for this fragment are given by $\sum_{j=1}^{r} e_j$, where 
each $e_j$ can take values $\pm 1$. 

The second example we consider is a fragment whose IS is made of $L-4$ $0$'s. The configuration will have a single substring $1111$ or a single
substring $101$ in a sea 
of zeroes; the total number of such states is $2L-5$. We can think of these
as the states of
a particle which can be either in a state $a_{j+3/2} = 1111$ at sites $(j,j+1, j+2,j+3)$, where $j=1,2,\cdots,L-3$, or in a state $b_j = 101$
at sites $(j-1,j,j+1)$, where $j=2,L-1$. The Hamiltonian can take the state $| a_{j+3/2} \ra$ to either $| b_{j+1} \ra$ or $| b_{j+2} \ra$. This gives us
a tight-binding model of a particle that moves on a finite line with $2L-5$
sites. We then find that the energy levels of this effective Hamiltonian are 
\begin{equation} E_k = 2 \cos [\pi k/(2L-4)], {\rm ~ where~} k=1,\cdots,2L-5.
\end{equation}
\vspace*{.2cm}

\subsection{Description of the largest fragment}
\label{sec3e}

We now consider the largest fragment, which includes all the states reachable from the configuration of all $1$'s. The corresponding IS reduces to one of the four possibilities $\phi$ (null 
string) or $1$ or $11$ or $111$. 
Let $D_L$ denote the size of the fragment corresponding to the
IS given by $111 \cdots 111$, i.e., 1 repeated $L$ times. We will compute the
generating function
\beq V(x) ~=~ \sum_{L=0}^\infty ~D_L ~x^L. \label{vx} \eeq
Following a lengthy calculation whose details are shown in Appendix~\ref{appA},
we obtain the expression 
\bea V(x)=\frac{1}{1 ~-~ x ~-~ \left(\frac{1 ~-~ \sqrt{1-12x^4}}{6x}\right)}. \label{Vexp} \eea
Writing this in the form given in Eq.~\eqref{vx}, we find that
the growth of $D_L$ for large $L$ is determined by the singularities of $V(x)$ 
lying closest to the origin~\cite{HariMenon_1995,menon_1997,Barma_1994}.
According to Eq.~\eqref{Vexp}, these singularities lie
at $x_{0}^4= 1/12$, namely, the fourth roots of $1/12$. Hence $D_{L}$ grows 
as $(1/|x_0|)^{L}$, i.e., $1.8612^{L}$ for large $L$. To confirm
this, we Taylor expand Eq.~\eqref{Vexp} which generates the series
\begin{widetext}
\bea
V(x)&=&1+x+x^2+2x^3+3x^4+4x^5+6x^6+12x^7+19x^8+28x^9+46x^{10}+92x^{11}+150x^{12}
+232x^{13}+396x^{14}\non \\
&&+792x^{15}+1315x^{16}+2092x^{17}+3646x^{18}+7292x^{19}+12258x^{20}+19864x^{21}
+35076x^{22}+70152x^{23}+ \cdots, \non \\
&& \eea
\end{widetext}
We have checked that these numbers perfectly agree with those obtained by brute force numerical enumeration. 
We note that the exponential growth rate of the largest sector with $L$ is slower than the total number of states $2^L$, and goes to zero in the limit of large $L$, which establishes the strong 
fragmentation~\cite{sala_ergo_2020} of the Hilbert space in this model.

\subsection{Typical and atypical fragments}
\label{typical}

Having discussed the idea of IS and seen several examples of fragments, it is
useful to discuss {\it typical} and {\it atypical} fragments. It is difficult to 
differentiate between the two very precisely but we may proceed roughly as
follows.

\begin{figure}[h!]
\includegraphics[width=\columnwidth]{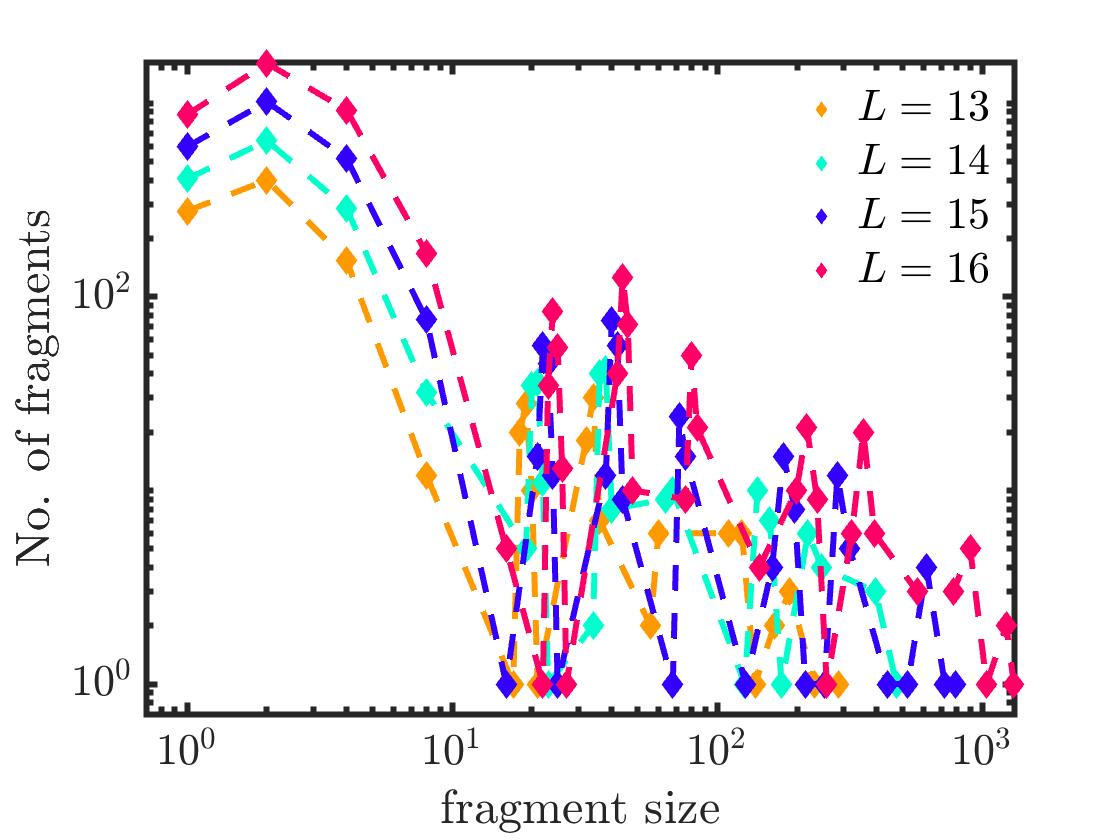}%
\caption{Plot showing the distribution of number of fragments versus sizes of fragments for four different system sizes, $L=13,14,15$ and $16$, with OBC.}
\label{secdis} \end{figure}

We first show the distribution of fragment numbers for different system sizes 
with OBC in Fig.~\ref{secdis}. A fragment size equal to 1 (left edge of the figure) 
corresponds to frozen fragments which are discussed in Sec.~\ref{sec3c}. The 
largest fragment (right edge of the figure) will be discussed in Sec.~\ref{sec3e}.

We can qualitatively estimate the average size of a fragment as follows.
While the dimension of the total Hilbert space is $2^L$, we saw earlier that
the number of fragments
grows asymptotically as $\tau^L$. We may therefore expect a typical fragment to have 
a size of the order of $(2/\tau)^L \simeq 1.236^L$. For $L=13, ~14, ~15$ and $16$, 
this goes from about $16$ to $30$. Indeed we see that $16 - 30$ lies near the 
middle of the ranges of fragment sizes shown in Fig.~\ref{secdis}.

\begin{figure}[h!]
\includegraphics[width=\columnwidth]{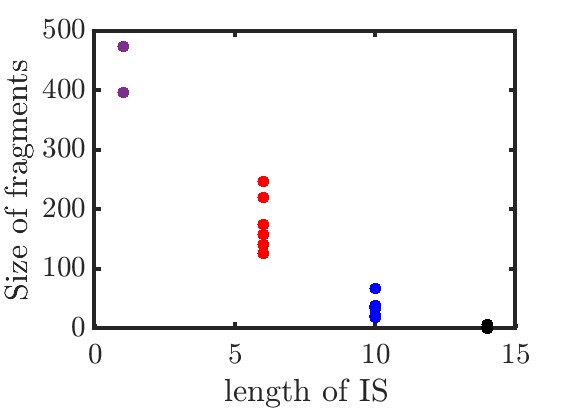}%
\caption{Plot of the sizes of fragments versus the lengths of their IS for 
$L=14$ with OBC.} \label{ISsize} \end{figure}

Next, we examine numerically if there is any relation between the size of a
fragment and the length of its IS. This is shown in Fig.~\ref{ISsize} for $L=14$, and we 
observe that the two quantities appear to be inversely related. (Note that the IS lengths can only be equal to 
$L, ~L-4, ~L-8, ~L-12, ~\cdots$). We see from the figure that two fragments with the 
same IS lengths may have different sizes, but 
the ratio of their sizes is of order $1$. Frozen fragments have only one state 
each and the length of their IS is equal to $L$; hence they correspond to the
point at the bottom right of Fig.~\ref{ISsize}.

We may now define an atypical fragment as one which has an exponentially
large number of states (in the limit $L \to \infty$), and whose IS has a length $l$
where $L-l \ll L$. With this definition, both frozen fragments (which have $l=L$)
and the simple integrable fragments discussed in Sec.~\ref{simple} are atypical.
On the other hand, the largest fragment and other fragments with short IS are
all typical. The significance of a typical fragment is that it is expected to
behave like a standard macroscopic ensemble, and we may expect a weak 
version of ETH to hold within it.

We have found numerically that the ground state and low-lying excited states 
generally lie in the largest fragment and other fragments which have comparable 
numbers of states. Thus the low-energy properties of this model are dominated by the
largest fragments.

\section{Subspace-restricted ETH in systems with strong Hilbert space fragmentation}
\label{sec4}

The existence of atypical fragments, like the frozen states and integrable 
fragments (for example, fragments with $n$ blinkers), implies that strong ETH is
not satisfied with respect to the full Hilbert space in our model.
This motivates us to modify the ETH as follows. Suppose
that a Hamiltonian in a basis given by products of local states has a fragmented
structure, such that energy eigenstates have non-zero components only within a
single fragment. Then it seems natural to postulate that in each fragment,
correlation functions of local observables calculated in the eigenstates of the Hamiltonian will tend to values corresponding to a restricted microcanonical 
ensemble, in which all the the eigenstates (except for a set of measure zero) 
having a given energy density are equally likely. 
We will call this a weak subspace-restricted ETH.
In our model, this should be satisfied in all the typical fragments but it need 
not be satisfied in atypical fragments.

\begin{figure}[htb]
\includegraphics[width=\columnwidth]{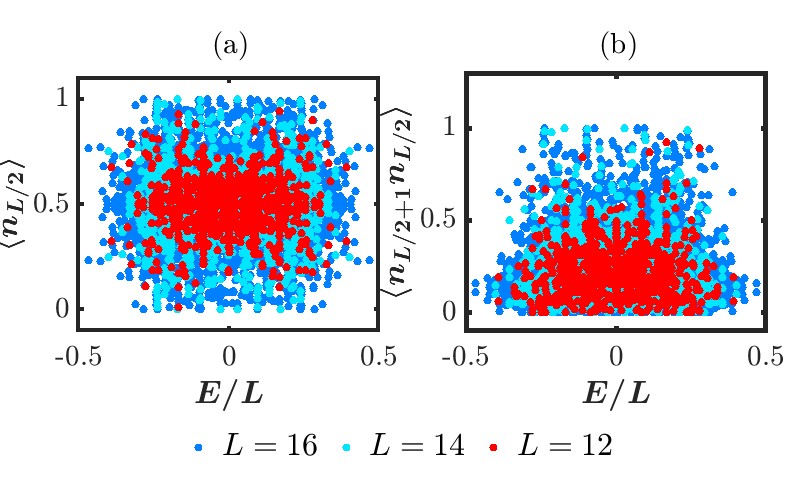}\\
\includegraphics[width=\columnwidth]{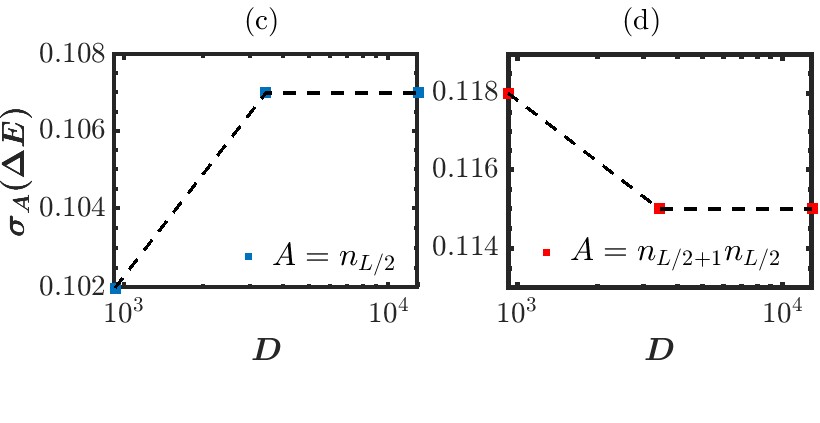}
\caption{Scatter plots of (a) $\langle n_{L/2}\rangle$ and (b) 
$\langle n_{L/2+1}n_{L/2}\rangle$ for all the eigenstates as a function of 
energy density $E/L$ for different system sizes, $L=12,14$ and $16$. In all three 
cases, the widths of the distribution do not appear to decrease significantly 
with increasing $L$. The standard deviation $\sigma_A$ of $\langle n_{L/2}\rangle$ 
and $\langle n_{L/2+1}n_{L/2}\rangle$ as a function of Hilbert space dimension $D$ 
in an energy window of $\Delta E/L~=~0.025$ are shown in
plots (c) and (d). The values of $\sigma_A$ do not decrease significantly with 
increasing Hilbert space dimensions similar to Figs. (a-b); this behavior 
is also not consistent with strong ETH.} \label{local observable} \end{figure}

We will now check the validity of a weak version of ETH within the largest 
fragment, as well as the validity of ETH in other sectors in the full Hilbert 
space at half-filling for the four-site model. We first examine 
the variation of expectation values 
of local observables for all eigenstates of the Hamiltonian and the variation 
of the half-chain entanglement entropy $S_{L/2}$ as a function of the energy 
$E$ without resolving the fragmentation structure. In 
Figs.~\ref{local observable} (a) and (b), we 
show the expectation values of the local observables $\langle n_{L/2}\rangle$ and 
$\langle n_{L/2+1}n_{L/2}\rangle$ 
as a function of the energy 
density $E/L$ for all the eigenstates for three different system sizes, $L=12, 14$ and $16$. We see that the widths of the distributions do not decrease significantly with 
increasing system sizes. Moreover, we also analyze the standard deviations of the differences of $\langle E | A |E\rangle - \langle A \rangle_{E}$, where 
$\ket{E}$ denotes an eigenstate of the Hamiltonian at energy $E$, and $\langle A \rangle_{E}$ is the microcanonical expectation value of $A$ at energy $E$ 
obtained by averaging over eigenstates within an energy window $\Delta E/L=0.025$~\cite{beugeling_2014}. For our case, we consider $A$ to be $n_{L/2}$ 
and $n_{L/2+1}n_{L/2}$, and then examine the standard deviation $\sigma_A$ as a function of 
the total Hilbert space dimension $D$ at half-filling, as shown in Figs. \ref{local observable} (c) and (d). In both cases, we observe that the values of $\sigma_A$ do 
not change notably with increasing values of $D$. This confirms that the strong version of diagonal ETH is not valid in our model within the full Hilbert space at 
half-filling. A similar behavior was seen earlier in other models showing HSF ~\cite{sredniki_1994,deutsch_1991}. 

\begin{figure}[htb]
\stackon{\includegraphics[width=0.47\linewidth,height=3.5cm]{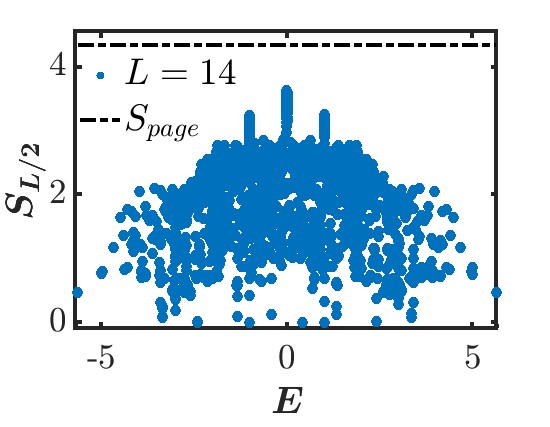}}{(a)}\stackon{\includegraphics[width=0.5\linewidth,height=3.5cm]{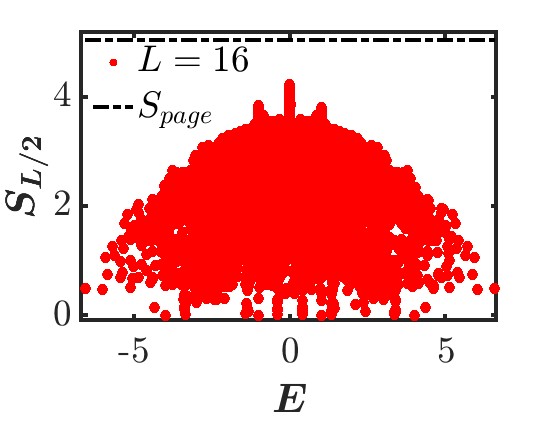}}{(b)}\\\stackon{\includegraphics[width=0.48\linewidth,height=3.5cm]{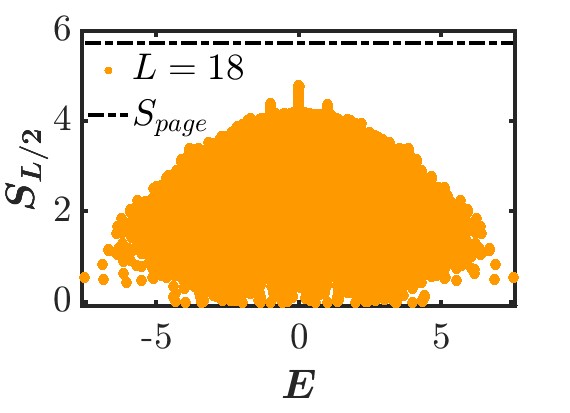}}{(c)}\stackon{\includegraphics[width=0.5\linewidth,height=3.5cm]{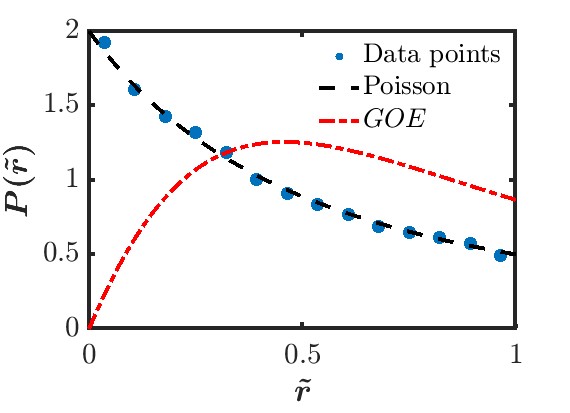}}{(d)}
\caption{(a-c) Plots showing the half-chain entanglement entropy $S_{L/2}$
versus $E$ for the full Hilbert space at half-filling for $L=14$, $16$ and $18$,
respectively, with OBC. In all three cases, the 
entanglement entropies of all the eigenstates do not lie within a narrow band,
and they have values much smaller than the thermal value shown by the dash-dot
lines at the top. Also, the widths of the spectra do
not decrease with increasing $L$, indicating that strong ETH is not valid. 
(d) Plot showing the probability distribution of $\tilde{r}$ for the sorted 
eigenspectrum for the full Hilbert space at half-filling for a $18$-site
system with OBC. (In order to break any discrete symmetries, we add a small 
uniformly distributed random on-site potential with strength $w=0.01$. The disorder 
preserves the fragment structure of the full Hilbert space). The distribution 
$P (\tilde{r})$ follows a Poison curve with $\langle r \rangle \simeq 0.383$.}
\label{ent_original} \end{figure}

Next, we discuss the spectrum of the half-chain 
entanglement entropy $S_{L/2}$ as a function of the energy $E$ for the full Hilbert space (without resolving the individual fragment) at half-filling with OBC
for three different system sizes, $L=14,16$ and $18$. This is shown in 
Figs. \ref{ent_original} (a-c). In all three figures, we see that many 
low-entanglement states are present in the middle of the spectrum. Further, the values of entanglement entropies for all the eigenstates are much 
smaller than the thermal value $S_{page} = (L \ln 2 - 1)/2$~\cite{page_1993} 
shown by the dash-dot lines; this again shows that ETH with respect to 
the full Hilbert space is not satisfied. Moreover, the entropies of the 
eigenstates do not lie within a narrow band when plotted against the energy.
Rather they are distributed over a wide range of values which are all
much smaller than the thermal value; this is exactly the opposite of what is typically observed for a system obeying strong ETH. Also, the width of the 
entanglement entropy spectrum does not shrink with increasing $L$ unlike a thermal system, indicating a manifestly non-thermalizing behavior as a 
consequence of strong HSF.

We next study the energy level spacing distribution which is often studied to 
probe whether a model is integrable or non-integrable~\cite{Wigner_1955,serbyn_2016,pal_2010}. It is 
well-known from the theory of random matrices that non-integrable systems described by random matrices show level repulsion, but integrable models, or models with extra 
conserved quantities do not. To quantify the
degree of level repulsion in a model, it is often better to study the level 
spacing ratios rather
than the level spacing distribution of the sorted eigenspectrum.
To do so, 
we define the level spacing ratios of the sorted eigenspectrum by $r_{n} = \delta_{n+1} / \delta_{n}$, where $\delta_{n}=E_{n+1}-E_{n}$ and $E_{n}$ is 
the $n$-th energy eigenvalue. If a system is integrable, $r$ follows the Poisson distribution, i.e., $P(r)=1/(1+r)^2$, while non-integrable 
systems described by real Hermitian Hamiltonians follow the Gaussian orthogonal ensemble (GOE) distribution~\cite{atas_2013,berry_1977} which is given by
the distribution
\bea P(r) ~=~ \frac{27}{8}\frac{(r+r^2)}{(1+r+r^2)^{5/2}}. \eea
It is also convenient to study the distribution of $\Tilde{r}$, which is defined as
\beq \tilde{r}_{n} ~=~ \frac{\text{min}(\delta_{n+1},\delta_n)}{\text{max}(\delta_{n+1},\delta_{n})} ~=~ \text{min}(r_{n},1/r_{n}). \eeq
For the two classes mentioned above, the distribution of $\Tilde{r}$ follows $P(\Tilde{r})=2P(r)\,\theta(1-r)$, where $\langle\tilde{r}\rangle=0.386$ for
Poisson and $0.536$ for GOE. 

To numerically compute the level spacing statistics of the 
consecutive energy levels for the sorted spectrum of the full system, we add 
a small uniformly distributed random on-site potential with strength $w=0.01$ for
an $18$-site system of $L=18$ with OBC to break all the discrete symmetries and to
eliminate any accidental degeneracies~\cite{moudgalya_PRX_2022,
Moudgalya_pairhopping_2020,Herviou_2021}. We note that
the presence of an on-site disorder preserves the fragmentation structure of the 
full Hilbert space. In addition, we also impose the half-filling condition 
to choose a particular $C_{1}$ symmetry sector, but we do not restrict the analysis 
within particular $C_{2}$ and $C_{3}$ symmetry fragments since these two global symmetries are only well-defined for $L=4n$. As shown in Fig. \ref{ent_original} 
(d), we see that the distribution of $\tilde{r}$, called $P(\tilde{r})$, is indistinguishable from a Poisson curve with $\langle \tilde{r}\rangle \simeq 0.383$;
this is very close to the value observed for an integrable system. The exponentially large number 
of dynamically disconnected fragments act as large number of conserved 
quantum numbers that forbid level repulsion as in an integrable system~\cite{Herviou_2021,kwan2023minimal}.

\begin{figure*}[htb]
\stackon{\includegraphics[width=0.33\linewidth]{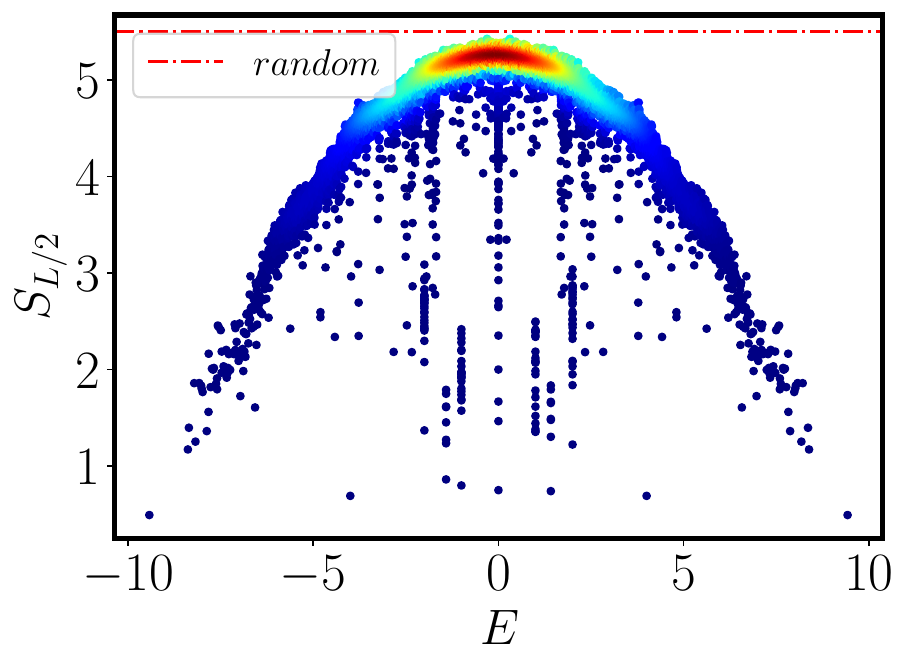}}{(a)}
\stackon{\includegraphics[width=0.33\linewidth]{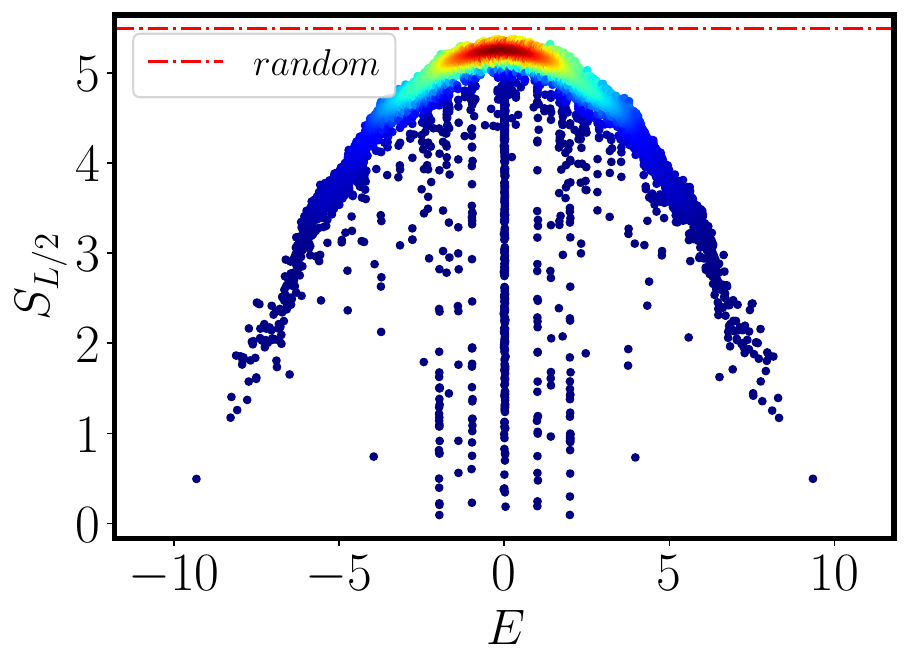}}{(b)}\stackon{\includegraphics[width=0.33\linewidth]{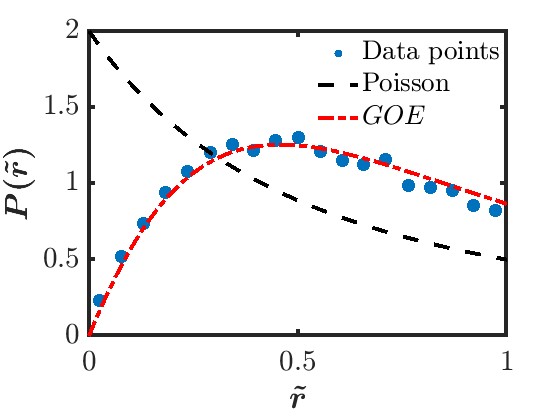}}{(c)}
\caption{(a-b) Plots showing the half-chain entanglement entropy $S_{L/2}$ versus $E$ for the largest fragment produced by enumerating the root configuration $1010 
\cdots 1010$ in the four-site model for $L=22$ (the fragment size is $19864$) 
with OBC with (a) no disorder and (b) a small uniformly distributed disorder with strength $w=0.01$, respectively. (a) The entropies for most of the eigenstates 
within this fragment lie within a narrow band, apart from 
some low-entanglement outlying eigenstates in the middle of the spectrum. (b) The spectrum in the presence of disorder exhibits identical features. In both 
cases, the red dash-dot line denotes the entanglement entropy of a random state within this particular fragment. (c) Plot of the probability 
distribution $P(\Tilde{r})$ versus $\Tilde{r}$ for case (b). $P (\Tilde{r})$ 
obeys the GOE distribution, indicating non-integrability of this fragment.}
\label{ent_fragment} \end{figure*}

Finally, we study the behavior of the largest fragment generated by enumerating the root state $1010 \cdots 1010$ for the original model with terms involving four
consecutive sites for $L=22$ with OBC (the size of this fragment is $19864$). In the three-site model, this state reduces to the state $111 \cdots 111$, 
which generates the largest fragment.
In Fig. \ref{ent_fragment} (a), we show the entanglement entropy 
as a function of $E$ for the largest fragment. We see that the entropies of 
most of the eigenstates fall on a curve, with a small number of 
outlying low-entanglement eigenstates in the middle of the spectrum. In Fig. \ref{ent_fragment} (b), we perform the same analysis with a small uniformly 
distributed random on-site disorder with strength $w=0.01$ to discard any discrete symmetries and to avoid any accidental 
degeneracies~\cite{moudgalya_PRX_2022,Moudgalya_pairhopping_2020}. 
(The random on-site disorder preserves the fragment structure). We see that the spectrum in this case shows features identical to the previous case; further, it
stabilizes the low-entanglement eigenstates, as can be seen in Fig. \ref{ent_fragment} (b). We find that the consecutive energy level spacing statistics for the eigenstates 
within this fragment with small disorder is consistent with GOE level statistics with 
$\langle \Tilde{r} \rangle \simeq 0.524$ 
which is close to the GOE value. This points towards non-integrability of the largest fragment~\cite{Moudgalya_pairhopping_2020}, as shown in Fig. \ref{ent_fragment} (c). 

\begin{figure}[htb]
\includegraphics[width=\columnwidth]{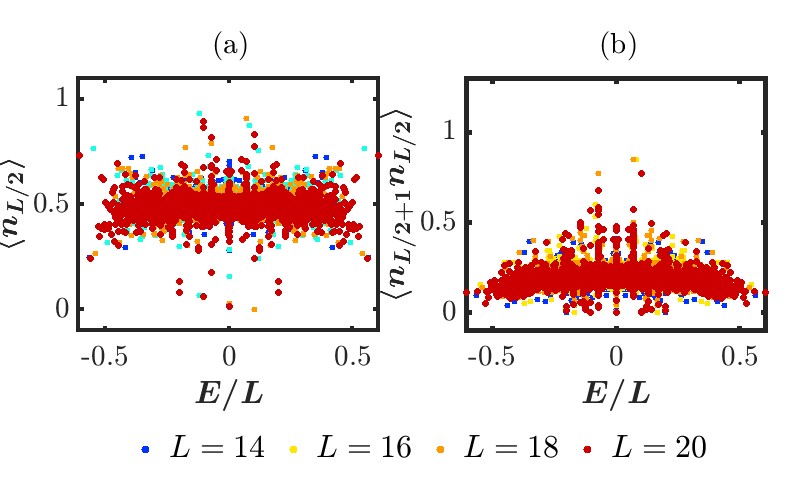}
\includegraphics[width=\columnwidth]{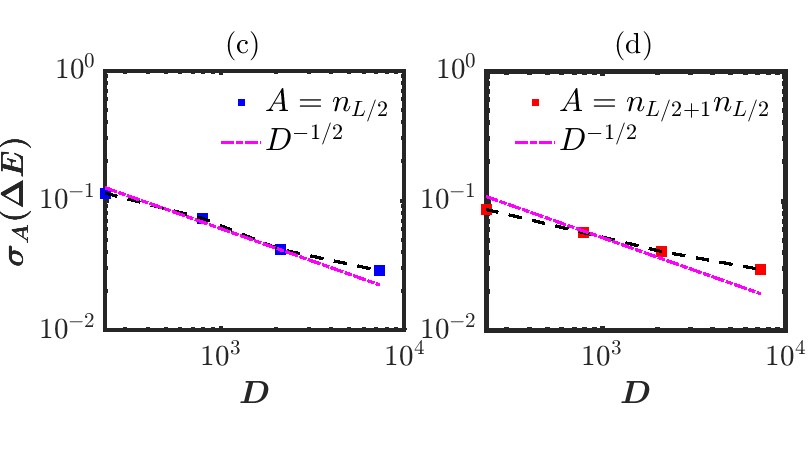}
\caption{Plots of (a) $\langle n_{L/2}\rangle$ and (b) 
$\langle n_{L/2+1}n_{L/2} \rangle$ for all the eigenstates 
within the largest fragment of the full Hilbert space. Most of the eigenstates within this fragment show a thermal behavior. In both cases, the bulk of the 
distribution becomes narrower with increasing $L$, but with some outlying states 
which do not fit within a narrow window of values of $\langle n_{L/2} 
\rangle$ and $\langle n_{L/2+1}n_{L/2} \rangle$. The standard deviations $\sigma_A$
of $\langle n_{L/2}\rangle$ and $\langle n_{L/2+1}n_{L/2} \rangle$ as a function of the size $D$ of the largest fragment within an energy window of $\Delta E/L~=~0.025$ are shown in
plots (c) and (d). We see that the values of $\sigma_A$ decrease with increasing sizes of largest fragment, which approximately scale as $1/\sqrt{D}$ with a slight 
deviation. The decreasing values of $\sigma_A$ with increasing fragment size supports the subspace-restricted ETH within the largest fragment.} \label{ETHsub} \end{figure}

In Figs.~\ref{ETHsub} (a) and (b), we show the average values of $n_{L/2}$ and
$n_{L/2+1} n_{L/2}$ for all the eigenstates within this subspace for $L=14,16,18$ 
and $20$ where the fragment sizes are $232, ~792, ~2092$ and $7292$, respectively. We see that
the width of this distribution becomes narrower with increasing system size. We then perform the same analysis as Figs. \ref{local observable} (c) and (d) within the 
largest fragment in Figs. \ref{ETHsub} (c-d). We see that the standard deviations
$\sigma_A$ of local observables decrease with increasing values of fragment size $D$. Moreover, $\sigma_A$ approximately scales as $~1/\sqrt{D}$ with some deviation, which 
has been seen earlier in the ETH obeying systems~\cite{beugeling_2014}. This behavior again indicates that the subspace-restricted diagonal ETH is satisfied within this 
fragment. However, there are some outlying states which do not show thermal behavior as shown in \ref{ETHsub} (a) and (b). 

\begin{figure}[htb]
\includegraphics[width=\columnwidth]{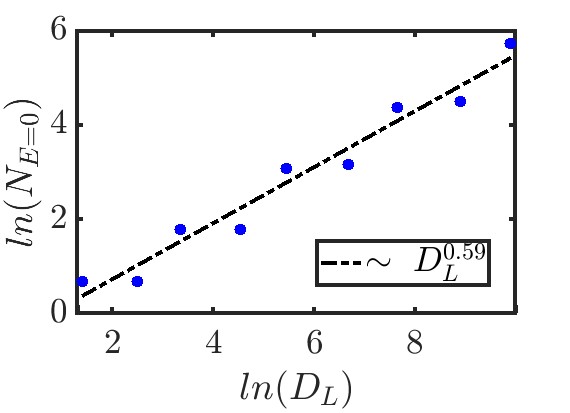}
\caption{Log-log plot of the number of zero energy states $N_{E=0}$ versus the
size $D_L$ of the largest fragment for system size $L$. The numerical fitting 
shows that $N_{E=0}$ grows as $D_{L}^{0.59}$.} \label{zero_energy_fit} \end{figure}

We also observe numerically that the largest fragment contains a large number of
states, $N_{E=0}$, with exactly zero energy. The variation of $N_{E=0}$ with the 
size $D_L$ of the largest fragment ($L$ is the system size) is presented in 
Table~\ref{zeroenergy}, and Fig. \ref{zero_energy_fit} shows a log-log plot
of $N_{E=0}$ versus $D_{L}$. The numerical fitting indicates that $N_{E=0}$ grows 
as $D_{L}^{0.59}$. It has been shown earlier in some 
models~\cite{schecter_2018,adithi_2023} that there are index theorems which 
give a lower bound on the growth of $N_{E=0}$ versus $D_L$.
In these models, index theorems imply that $N_{E=0} \gtrsim \sqrt{D_{L}}$. 
Fig.~\ref{zero_energy_fit} implies that
such a square root bound is also satisfied by the largest fragment in our 
four-site model. However we have not been able to derive this bound analytically
due to the lack of a simple structure of the states and of the Hamiltonian
within this fragment. Finally, we note that the square root bound also holds 
for the full Hilbert space. This is because the frozen states are trivially 
zero-energy eigenstates of the Hamiltonian, and we saw in Sec.~\ref{sec3c}
that the number of such states grows asymptotically as $1.466^{L}$. This is
much larger than a square root bound since $\sqrt{2^L} \simeq 1.414^{L}$.

We thus see that the largest fragment contains a large number of zero
energy states, and many of these have an anomalously
low entanglement entropy as shown in Fig.~\ref{ent_fragment} (a). The 
presence of these exceptional states implies that the subspace-restricted 
ETH is satisfied with the largest fragment in a weaker sense.

\begin{table}[htb]
\begin{tabular}{|c|c|c|} 
\hline
 $L$& $D_{L}$ & $N_{E=0}$\\
\hline
6&4&2 \\
\hline
8&12 &2\\
\hline
10 &28&6\\
\hline
12&92&6\\
\hline
14&232&22\\
\hline
16&792&24\\
\hline
18&2092&80\\
\hline
20&7292&90\\
\hline
22&19864&308\\

\hline
\end{tabular}
\caption{Table showing the system size $L$, the size $D_L$ of the largest 
fragment generated from the root state $1010....1010$ in the four-site model, 
and the number of zero energy states $N_{E=0}$ in this fragment.}
\label{zeroenergy} \end{table}

\begin{figure}[htb]
\includegraphics[width=\columnwidth]{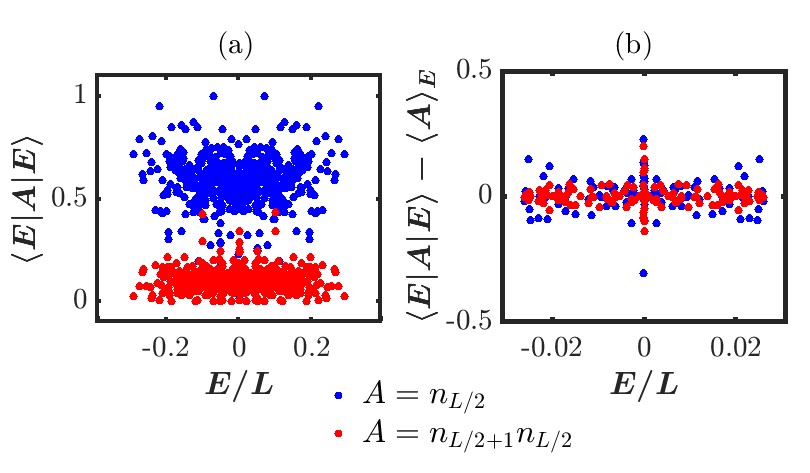}
\caption{(a) Expectation values of two few-body observables $A=n_{L/2}$ and $n_{L/2+1}n_{L/2}$ for all the eigenstates within the fragment generated from a root state $10100010010010101010$ for $L=20$ in the four-site model. (b) Plot showing the difference between the expectation values of the observables for an eigenstate at energy $E$ and their microcanonical value obtained by averaging over eigenstates within an energy window of $\Delta E=0.025$. The analysis shown in Fig. (b) has been performed for 
one-fifth of the eigenstates in the middle of the spectrum.}
\label{typifrag} 
\end{figure}

In Figs.~\ref{typifrag} (a-b), we study the thermal behavior of eigenstates within a fragment generated from a randomly chosen root state $10100010010010101010$ for $L=20$ 
in the four-site model; this fragment has dimension $D=574$. This root state reduces to $1110011011011111111$ due to bond-site mapping, whose IS, $11100110110$ has a length 
$(L+3)/2$ in the three-site model, unlike the largest fragment which has the shortest
IS. In Fig. \ref{typifrag} (a), we show the expectation values of two local observables 
$n_{L/2}$ and $n_{L/2+1}n_{L/2}$ for all eigenstates within this fragment. In Fig. \ref{typifrag} (b), we show the difference between the expectation values of the same 
two observables $\langle E|A|E\rangle$ for an eigenstate at energy $E$ from their microcanonical expectation value $\langle A\rangle_{E}$ 
obtained by averaging over eigenstates within an energy window $\Delta E=0.025$ about $E$. Moreover, we consider one-fifth of the total eigenstates with energies lying in 
the middle of the spectrum while doing this analysis. As this difference is centered around zero, it implies that each eigenstate satisfies diagonal ETH within this 
fragment. One can therefore conclude that all typical fragments satisfy subsector-restricted thermalization in our model in the thermodynamic limit.

Despite the fact that ETH is not valid with respect to the full Hilbert space as shown in Figs. \ref{local observable} and \ref{ent_original}, expectation values of local 
observables of eigenstates within sufficiently large fragments still satisfy ETH as we see in Figs. \ref{ent_fragment} and \ref{ETHsub}. Therefore, sufficiently large 
fragments still satisfy ETH even in the case of strongly fragmented systems.
This is also dubbed as Krylov-restricted thermalization in the literature~\cite{Moudgalya_pairhopping_2020,Moudgalya_review_2022,stephen_2024}. This 
kind of restricted thermalization has a significant impact on the dynamics of the 
system, in particular, an atypical dynamical behavior of correlation functions, which we 
will discuss in the next section.

\section{Dynamical signatures of Hilbert space fragmentation}
\label{sec5}

In this section, we will study autocorrelation functions of the fermion 
number at different sites of the four-site model. We will see that these provide
dynamical signatures of the absence of thermalization due to HSF.

\subsection{Long-time behavior of autocorrelation functions}

As a signature of the lack of thermalization due to strong HSF, we first investigate the behavior of the time-dependent correlation function 
$$C_{j}(t)=\langle \psi |(n_{j}(t)-1/2) (n_{j}(0)-1/2) |\psi\rangle,$$ 
where $n_{j}$ is the fermion number operator at site $j$, $\ket{\psi}$ being a typical random initial state in the full Hilbert space, which is chosen to have the form $\ket{\psi}=\sum_{j} a_{j} \ket{f_{j}}$ with $\sum_{j} |a_{j}|^2=1$, where the
$a_{j}$'s are random numbers and $\ket{f_{j}}$ denote
Fock space basis states. We will consider the case of half-filling and open 
boundary conditions.

\begin{figure}[h!]
\stackon{\includegraphics[width=0.48\linewidth]{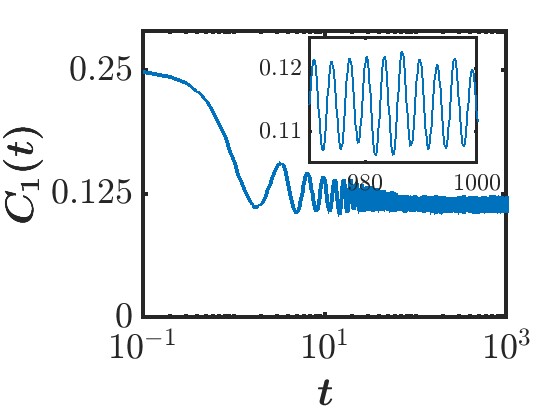}}{(a)}
\stackon{\includegraphics[width=0.48\linewidth]{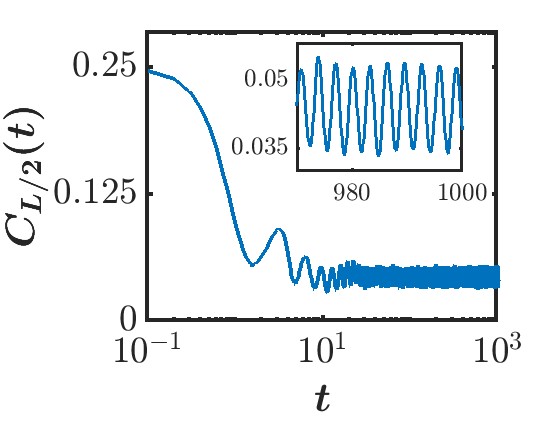}}{(b)}\\\stackon{\includegraphics[width=0.48\linewidth]{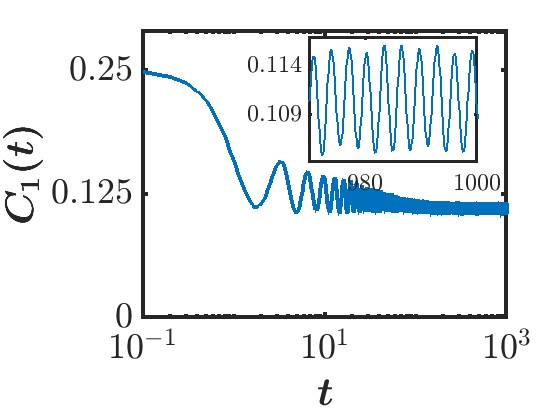}}{(c)}
\stackon{\includegraphics[width=0.5\linewidth]{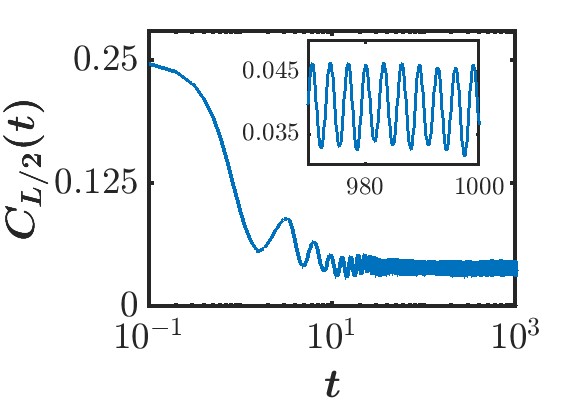}}{(d)}
\caption{(a-b) Plots showing the long-time behaviors of the boundary and bulk correlators, $C_{1}(t)$ and $C_{L/2}(t)$, respectively, starting from a typical random initial state at half-filling with OBC for
$L=18$. (c-d) Similar plots for $L=20$. (a) $C_{1}(t)$ oscillates around a finite saturation value of about $0.115$ at long times, revealing non-thermal 
behavior near the boundary of the system. (b) $C_{L/2}(t)$ 
oscillates around a finite saturation value of about $0.045$, which
is much smaller than the boundary case. (c) $C_{1}(t)$ for this case oscillates around a finite saturation of about $0.113$ at long times, which is slightly smaller compared 
to that observed in Fig. (a). (d) In this case, $C_{L/2}$ again shows similar behavior as Fig. (b), but it oscillates around a saturation value ($\sim0.04$), again slightly 
smaller compared to Fig. (b). In all four cases, the last ten oscillations for $970 
\leq t \leq 1000$ are shown in the insets of the plots.}
\label{autocorr1} \end{figure}

In thermal equilibrium, the autocorrelation 
function is expected to decay to zero as $1/L$ for a system of length $L$. In Fig. \ref{autocorr1} (a), we study the boundary 
autocorrelator, $C_{1}(t)$ for a system size $L=18$ with OBC at 
half-filling for a random initial state in the Hilbert space. We find
persistent oscillations around a finite saturation value of about $0.115$ (shown in the inset of the plot) up to
a long time $t \sim 10^3$. We then study the same function in the middle of
the system, $C_{L/2}(t)$, for the same system size in Fig. \ref{autocorr1} (b). We observe that $C_{L/2}(t)$ saturates to a much smaller value of about $0.045$ (shown 
inset of the plot) at long times. In a similar manner, we show the same 
quantities for $L=20$ in Fig. \ref{autocorr1} (c-d). We see that the behaviors of $C_{1}(t)$ and $C_{L/2}(t)$, including the period of oscillations, do not 
significantly change with increasing system size. However, both quantities oscillate around finite saturation values given by $0.113$ and $0.04$, respectively (shown in 
the insets of the plots), which are slightly smaller compared to Figs. \ref{autocorr1} (a) and (b), respectively. 
We therefore conclude that the boundary correlator behaves in a different 
manner from the bulk correlator.

\begin{figure}[htb]
\includegraphics[width=\columnwidth]{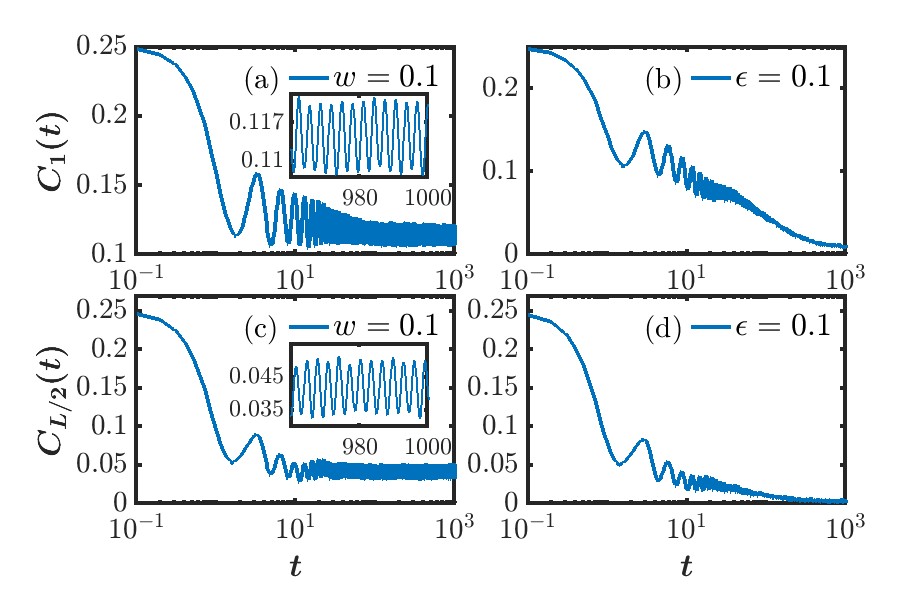}%
\caption{(a-b) Plots showing the long-time behavior of the boundary correlator $C_{1}(t)$ starting from a typical random initial state for $L=18$ at 
half-filling with OBC in the presence of two kinds of perturbations, uniformly distributed random disorder with disorder strength $w=0.1$, and the perturbation 
term with $\epsilon=0.1$ shown in Eq.~\eqref{hp}.
(a) $C_{1}(t)$ oscillates around a finite value of about $0.12$ in the 
long-time limit just as in the unperturbed case. (b) $C_{1}
(t)$ decays to zero at long times after exhibiting a non-trivial intermediate-time dynamics close to the saturation value of the unperturbed model.
(c-d) Plots showing the long-time behaviors of bulk correlators, $C_{L/2}(t)$ 
starting from a random initial state again for the same two cases. (c) $C_{L/
2}(t)$ in the presence of random disorder again shows a behavior similar to the unperturbed case. (d) $C_{L/2}(t)$ for this case decays to zero at long times, 
after exhibiting non-trivial intermediate-time dynamics like the boundary case. In Figs. (a) and (c), the last ten oscillations for $970 \leq t \leq 1000$ are shown in the insets of the plots.}
\label{autocorr2} \end{figure}

Further, strong HSF leads to an non-uniform profile of correlation function near the edge of the chain in our model
model as was observed earlier in this context for other models~\cite{sala_ergo_2020,sala_stat_loc_2020,moudgalya_PRX_2022}. We 
examine the robustness of this non-uniform profile near the edge against perturbations by including two types of terms in the four-site Hamiltonian. The first one is a
uniformly distributed random on-site potential of strength $w=0.1$, and the
results are shown in Figs. \ref{autocorr2} (a) and (c). The second
one involves the Hamiltonian
\beq H' ~=~ \sum_{j}\left[ (n_{j-1}-n_{j+2})^{2} ~-~ \epsilon \right] (c_{j}^{\dagger}c_{j+1}+c_{j+1}^\dagger c_j) \label{hp} \eeq 
where $\epsilon=0.1$, giving the results shown in Figs. \ref{autocorr2} (b) and 
(d). The first one breaks all the discrete symmetries present in 
the four-site model, but it 
preserves the fragmentation structure of the full Hilbert space. On the other hand, the second one preserves all the discrete symmetries but modifies the 
fragmentation structure of the full Hilbert space; the numbers of fragments 
generally decreases for this case since the $\epsilon$ terms connects certain
states which are not connected otherwise. As shown in Figs. \ref{autocorr2} (a) and (c), the long-time behaviors of $C_{1}$ and $C_{L/2}$ in the first case remains the same as in the unperturbed case, exhibiting an absence 
of thermalization. For the second case, both the correlators decay to zero after showing a non-trivial intermediate-time dynamics, as can be seen in Figs. 
\ref{autocorr2} (b) and (d). 

We will now explain the long-time saturation value of the autocorrelation functions by taking the fragmentation structure of the full Hilbert space into account. It can be 
shown that the equilibrium value of $C_{j}(t)$ predicted by the ETH hypothesis is zero for all
values of $j$ if our model thermalizes. Therefore, the nonuniform profile of autocorrelation function near the edge of the chain shown in Fig. \ref{autocorr1} 
is an atypical behavior which arises 
as a consequence of subspace-restricted ETH due to strong HSF. This behavior
can be explained with the help of the Mazur inequality~\cite{MAZUR1969533}, 
which applies to the long-time averages of autocorrelation functions in the
context of thermalization. 
For quantum systems, an exact Mazur-type equality was obtained by Suzuki~\cite{SUZUKI1971277}, which takes into account existence 
of constants of motion in the problem. In the same spirit, 
the value of the Mazur bound for fragmented Hilbert spaces is changed by taking 
into account the structures of invariant subspaces. We do this as follows.

We define $P_{i}$ as the projection operator onto a particular fragment $\mathcal{H}_{i}$ with size $\mathcal{D}_{i}$. The set of 
projectors onto different fragments form a complete orthogonal set of conserved quantities such that $P_{i}P_{j}=\delta_{ij}P_{j}$. Using these, we define the long-time 
averaged autocorrelation functions 
\bea F_j = && \lim_{T\rightarrow0}\,\frac{1}{T} ~\int_{0}^{T}\,dt\,
\langle( n_{j}(t)-\frac{1}{2}) (n_{j}(0)-\frac{1}{2}) \rangle, \non 
\eea
where $n_{j}$ is the fermion number operator at site $j$. These satisfy the inequality due to Mazur~\cite{MAZUR1969533}
\bea F_j \geq ~\sum_{i} ~\frac{ \left( {\rm Tr} [P_{i}(n_{j}-\frac{1}{2})
P_{i}] \right)^{2}}{D D_{i}} ~\equiv~ C_{j}(\infty), \label{Mazur} \eea
Here $D_i$ is the dimensionality of the $i$-th fragment, and $D$ is the total Hilbert space dimension (here $D = 2^N$).

\begin{figure}[htb]
\includegraphics[width=\columnwidth]{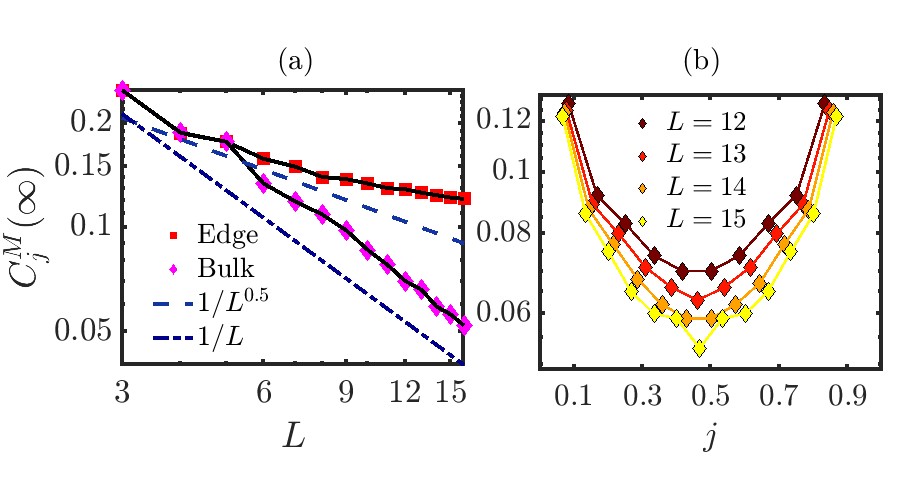}%
\caption{(a) Log-log plot showing the lower bounds of the infinite-time boundary and bulk correlators predicted by the Mazur inequality for different system sizes. In the 
bulk of the chain, the Mazur bound follows a $1/L$ curve instead of $1/\sqrt{L}$ (as depicted by two guiding lines), while near the edge of the chain, the Mazur bound 
shows a localized behavior and saturates to approximately $0.12$. (b) The variation of the Mazur bound across the chain for different sizes 
exhibits a non-uniform profile due to the strong HSF in our model.}
\label{Mazurbound} \end{figure}

In Fig. \ref{Mazurbound} (a), 
we plot the variation of the infinite-time saturation value of boundary and bulk 
autocorrelation functions, $C_{1}^{M}(\infty)$ and $C_{L/2}^{M}(\infty)$, obtained using Eq.~\eqref{Mazur} for an $L-$site system with OBC. We find that the Mazur 
bound in the bulk of the chain decays as $1/L$ for comparatively large system sizes like the
assisted pair-flipping model. On the other hand, the Mazur bound at the boundary of the chain
saturates to approximately $0.12$. We show the Mazur bound as a function of the site
index for different system sizes in Fig. \ref{Mazurbound} (b); this shows a non-uniform profile, being smaller at the centre compared to the ends. Further, HSF leads to 
localization close to the edge of the chain which has been dubbed as ``statistical 
edge localization"~\cite{sala_stat_loc_2020,Moudgalya_review_2022}; we see this 
in the long-time behavior of boundary autocorrelation functions shown in Figs. 
\ref{autocorr1} (a) and (b). We note here that the late-time average of the bulk autocorrelator decaying with system size as $1/L$ usually indicates thermal 
behavior of the bulk states. On the other hand, the localized profile of the 
autocorrelator near the edge shows non-thermal nature of the boundary spectrum. This implies that the non-local conserved quantities arising due to HSF do not 
have any significant impact on the thermal behavior of the bulk states for our 
model, in contrast to the boundary states. However, this numerical observation
requires a more careful investigation since our analysis has only been carried 
out for rather small system sizes. In some cases, it has been observed that such thermal behavior of the bulk autocorrelators is an artifact of limited system 
sizes, and the decay can deviate significantly from $1/L$ in the thermodynamic
limit. Specifically, the decay can go as $1/L^a$ where $a <1$~\cite{hart2023exact}.

\subsection{Dynamics of entanglement entropy}

\begin{figure}[h!]
\stackon{\includegraphics[width=0.514\linewidth]{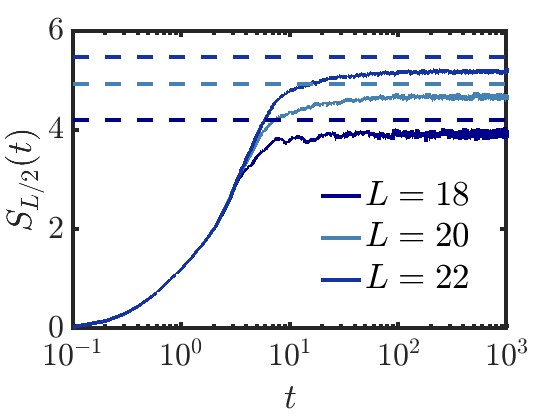}}{(a)}%
\stackon{\includegraphics[width=0.516\linewidth]{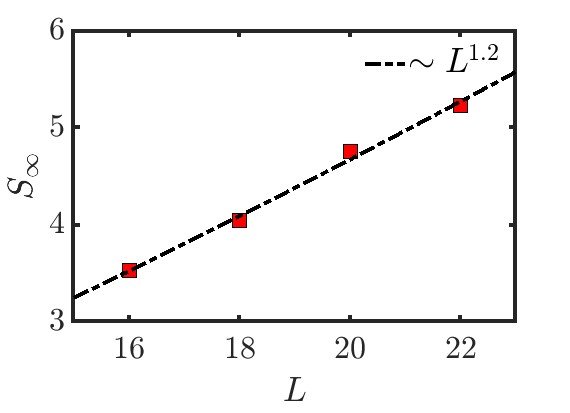}}{(b)}
\caption{(a-b) Growth of entanglement entropy with time for three different system sizes, $L=18,20$ and $22$ with OBC, starting from the Neel state, $101010 \cdots$. 
In all three cases, the entanglement entropy quickly saturates to a volume law as 
shown in Fig. (b), which is much smaller than the thermal value of the entropy 
for the full system, $S_{page}=(L \ln 2 - 1)/2$. The saturation 
values for all three cases are very close to the entanglement for a random 
initial state in the Hilbert space within largest fragment, depicted by the
three dashed lines in Fig. (a).} \label{entanglement_entropy} \end{figure}

To complement our previous findings, we study the dynamics of the entanglement
entropy starting from the Neel state $101010 \cdots$ for three different system 
sizes, $L=18,20$ and 
$22$ with OBC. This is shown in Fig. \ref{entanglement_entropy} (a). For all 
three cases, we see that the entanglement entropy quickly saturates to a 
volume law as shown in Fig. \ref{entanglement_entropy} (b). Moreover, the 
saturation value is much smaller than the thermal value of the entropy of the
full system, i.e., $S_{page}= (L \ln 2 - 1)/2$~\cite{page_1993}. 
The saturation value for all three cases are found to be quite close to the 
value of the entanglement entropy obtained 
for a random state on the full Hilbert space within the largest HSF sector, as depicted by the three dashed lines. These observations are in agreement with our 
previous findings, i.e., the largest fragment obeys a weaker form of 
subsector-restricted ETH~\cite{Moudgalya_pairhopping_2020}.

\section{Correlated-hopping model as the large interaction limit of a $t-V$ model}
\label{sec6}

In this section, we will show that our correlated-hopping model involving terms with four consecutive sites can be obtained by taking a particular large
interaction limit of a model of spinless fermions. 
We consider a model with a nearest-neighbor hopping which we will set equal to 1, and nearest-neighbor density-density interaction terms and on-site potentials 
which repeat with a periodicity of four sites.

We consider the Hamiltonian
\bea H_4 &=& \sum_{j} ~[~ c_{j}^{\dagger}c_{j+1} ~+~ c_{j+1}^\dagger c_j \non \\
&& ~~~~~~~~~+~ V_j ~(2n_{j}-1) ~(2n_{j+1}-1)~], \label{ham4} \eea
where $V_j$ varies with $j$ with period four, namely, $V_{4j+1} = V_1$, $V_{4j+2} = V_2$,
$V_{4j+3} = V_3$ and $V_{4j+4} = V_4$. 
We now consider various 
correlated-hopping processes involving four consecutive sites, namely, 
$1101 \leftrightarrow 1011$, $0100 \leftrightarrow 0010$, $1100 \leftrightarrow 1010$ and $0101 \leftrightarrow 0011$, for two different interaction patterns 
which are related to each other through translation by one site. We list 
the energy costs 
for the left and right sides of these correlated-hopping processes in the limit 
$V_{1}, ~V_{2}, ~V_{3}, ~V_{4} \rightarrow\infty$ in Table \ref{Table VII}.
We see from the table that the interaction energy costs will be equal
for the left hand and right hand sides of the processes in rows 3, 4, 6 and 7
(where the occupation numbers $n_j$ are unequal on the first and fourth sites) if $V_1 = - V_3 = V$ and $V_2 = - V_4 = V'$. Simultaneously, the interaction energies 
will {\it not} be equal for the left hand and right hand sides of the processes
in rows 1, 2, 5 and 6 (where the occupation numbers are equal on the first and
fourth sites) if $V, ~V' ~\ne~ 0$. Hence, in the limit $V \to \pm \infty$ and $V' \to \pm \infty$, hopping
between sites $j+1$ and $j+2$ is allowed if and only if $n_j \ne n_{j+3}$.
Note that $V$ and $V'$ can differ from each other in general.

\begin{widetext}
\begin{center}
\begin{table}[htb]
\begin{tabular}{|c|c|c|c|} 
\hline
Pattern of & Correlated-hopping & Energy cost for left hand side of 
& Energy cost for right hand side of \\
$V_j$ & process & the process in the second column & the process in the 
second column \\
\hline
$V_{1}\,V_{2}\,V_{3}\,V_{4}$&$1101 ~\leftrightarrow~ 1011$& $V_{1}-V_{2}-V_{3}$&$-~ 
V_{1}-V_{2}+V_{3}$\\
\hline
$V_{1}\,V_{2}\,V_{3}\,V_{4}$&$0100 ~\leftrightarrow ~0010$& $-~ V_{1}-
V_{2}+V_{3}$&$V_{1}-V_{2}-V_{3}$\\
\hline
$V_{1}\,V_{2}\,V_{3}\,V_{4}$&$1100 ~\leftrightarrow ~1010$& $V_{1}-V_{2}+V_{3}$&$-~ 
V_{1}-V_{2}-V_{3}$\\
\hline
$V_{1}\,V_{2}\,V_{3}\,V_{4}$&$0101 ~\leftrightarrow ~0011$& $-~ V_{1}-V_{2}-
V_{3}$&$V_{1}-V_{2}+V_{3}$\\
\hline
$V_{2}\,V_{3}\,V_{4}\,V_{1}$&$1101 ~\leftrightarrow ~1011$& $V_{2}-V_{3}-V_{4}$&$-~ 
V_{2}-V_{3}+V_{4}$\\
\hline
$V_{2}\,V_{3}\,V_{4}\,V_{1}$&$0100 ~\leftrightarrow ~0010$& $-~ V_{2}-
V_{3}+V_{4}$&$V_{2}-V_{3}-V_{4}$\\
\hline
$V_{2}\,V_{3}\,V_{4}\,V_{1}$&$1100 ~\leftrightarrow ~1010$& $V_{2}-V_{3}+V_{4}$&$-~ 
V_{2}-V_{3}-V_{4}$\\
\hline
$V_{2}\,V_{3}\,V_{4}\,V_{1}$&$0101 ~\leftrightarrow ~0011$& $-~ V_{2}-V_{3}-
V_{4}$&$V_{2}-V_{3}+V_{4}$\\
\hline
\end{tabular}
\caption{Energy costs arising from the $V_j$ term in Eq.~\eqref{ham4} for the 
left and right sides of various correlated-hopping processes for two sets of
patterns of $V_j$ related to each other through translation by 
one site in the limit $V_{1}$, $V_{2}$, $V_{3}$, $V_{4}\rightarrow\infty$.}
\label{Table VII}
\end{table}
\end{center}
\end{widetext}

Our analysis thus puts forward an experimentally realizable model which 
reduces to the four-site correlated-hopping model in the large interaction limit.
Namely, we have to consider a model for which the Hamiltonian is given by
Eq.~\eqref{ham4}, with
\bea V_{4j+1} &=& - ~V_{4j+3} ~=~ V, \non \\
{\rm and}~~~~ V_{4j+2} &=& - ~V_{4j+4} ~=~ V'. \label{finite-V Ham} \eea

Before ending this section we would like to point out that our correlated-hopping 
model can also emerge as an effective
Hamiltonian due to an interplay between dynamical localization, resonance and interactions
in a periodically driven system with an on-site potential with a spatial periodicity of four sites~\cite{aditya_2023}.

\section{Comparison with a different model showing Hilbert space fragmentation}
\label{sec7}

It is interesting to contrast various results for our model and a different model
showing HSF which has been studied extensively~\cite{tomasi_2019,
Moudgalya_review_2022,Moudgalya_pairhopping_2020,khemani_2020}. This is again 
a one-dimensional model with spinless fermions but with a Hamiltonian
\beq H_5 ~=~ \sum_{j} ~[1 ~-~ (n_{j}-n_{j+3})^{2}] \,(c_{j+1}^{\dagger}c_{j+2}
+ c_{j+2}^{\dagger}c_{j+1}). \label{ham5} \eeq
This Hamiltonian connects the following pairs of states involving four consecutive sites,
\bea 1101 &\leftrightarrow& 1011 \non\\
0100&\leftrightarrow &0010. \label{proori5} \eea
This comparison is particularly relevant for our study since this model is also a correlated-hopping model involving four consecutive sites just like ours. However, 
this model allows nearest-neighbor hoppings if the sites to the left and right of those two sites have equal particle numbers, unlike our model which enables 
nearest-neighbor hoppings if the sites to the left and right have different particle numbers. Defining $n_j = c_j^\dagger c_j$ as before, we find that
there are three global symmetries: total particle number $C_1$, and two other
quantities $C_{4}$ and $C_{5}$ given by
\bea C_{1} &=& \sum_j ~n_{j}, \non\\
C_{4} &=& \sum_j ~n_{2j} n_{2j+1}, \non\\
C_{5} &=& \sum_j ~n_{2j-1} n_{2j}. \label{globalsym5} \eea

As pointed out in Refs.~\onlinecite{lenart_2021}-\onlinecite{pozsgay_2021}, this 
model can be mapped to a model with a Hamiltonian
which involves three consecutive sites. On doing the mapping in Eq.~\eqref{2to1},
we obtain a model where only the following transitions are allowed,
\beq 110 ~\leftrightarrow~ 011. \label{trans2} \eeq
The Hamiltonian of this model is 
\beq H_6 ~=~ \sum_j n_{j+1} ~(d_j^\dagger d_{j+2} ~+~ d_{j+2}^\dagger d_j),
\label{ham6} \eeq
where $n_j = d_j^\dagger d_j$.
This Hamiltonian is number conserving, unlike the Hamiltonian in Eq.~\eqref{ham2}.
Note, however, that the total particle numbers $\sum_j n_j$ for the models in Eqs.~\eqref{ham5} 
and \eqref{ham6} are not related to each other in any simple way.

It turns out that a transition of the form given in Eq.~\eqref{trans2} was studied
many years ago in a classical model of diffusing dimers undergoing Markov 
evolution~\cite{menon_1997}. In that
work, a complete solution for the numbers and sizes of fragments was found. 
For large system sizes, it was shown that the number of fragments grows exponentially
as $\tau^L$. The number of frozen sectors is also found to grow as 
$\tau^L$,~\cite{aditya_2023} unlike our model
where it grows as $1.466^L$. Further, it was shown in Ref.~\onlinecite{menon_1997} that
for a system with OBC, the different fragments can be characterized uniquely by the 
numbers of three kinds of short strings, namely, $N_A$ strings given by $11$, $N_B$ 
given by $10$, and $N_C$ given by $0$. The number of states in a fragment 
$(N_A, N_B, N_C)$ was shown to be
\beq D_{N_A,N_B,N_C} ~=~ \frac{(N_A ~+~ N_B ~+~ N_C)!}{N_A! ~(N_B ~+~ N_C)!}. 
\label{nabc} \eeq
For a system with $L$ sites, we must have $2N_A + 2N_B + N_C = L$. The
filling fraction of particles is given by $(2N_A + N_B)/L$. The frozen fragments
with $D_{N_A,N_B,N_C} =1$ correspond to either $N_A = 0$ and $2N_B + N_C = L$, or $N_A = 
L/2$ and $N_B = N_c = 0$ (i.e., a string of $L$ $1$'s).

We can now find how the number of states in an arbitrary fragment grows with 
$L$. We define
\beq \al ~=~ \frac{N_A}{L}, ~~~~~~\beta ~=~ \frac{N_B}{L}, ~~~~~~\ga ~=~ \frac{N_C}{L}.
\eeq
These parameters satisfy $\al, ~\beta, ~\ga ~\ge~ 0$, and $2 \al + 2 \beta + \ga = 1$.
Eliminating $\beta$, we see that the parameters $(\al, \ga)$ lie in a triangular region
which is bounded by the lines $\al =0$, $\ga = 0$ and $2 \al + \ga = 1$ (where 
$\beta = 0$). We now consider the limit $L \to \infty$ holding $\al, ~\beta, ~\ga$ 
fixed. Using Eq.~\eqref{nabc} and Stirling's approximation, we find that the number 
of states grows as $\mu^L$, where $\mu$ is a function of $\al, ~\ga$ given by
\beq \mu (\al,\ga) ~=~ \frac{(\frac{1}{2} ~+~ \frac{\ga}{2})^{(1+\ga)/2}}{\al^\al 
~(\frac{1}{2} ~+~ \frac{\ga}{2} ~-~ \al)^{(1+\ga-2\al)/2}}. \label{mualga} \eeq
We thus see that $\mu (\al,\ga)$ varies continuously over the triangular region.
The minimum value of $\mu (\al,\ga)$ is equal to 1; this occurs on the line
$(\al = 0, 0 \le \ga \le 1)$, and at the point $(\al = 1/2,\ga = 0)$.
We will now find the maximum value of $\mu (\al,\ga)$.
A numerical search shows that $\mu (\al,\ga)$ attains its maximum on the line 
$2 \al + \ga = 1$. On that line, Eq.~\eqref{mualga} simplifies to
\beq \mu (\al) ~=~ \frac{(1 ~-~ \al)^{1-\al}}{\al^\al ~(1 ~-~ 2 \al)^{1-2\al}},\label{mualga2} \eeq
where $0 \le \al \le 1/2$. We find analytically that this has a maximum at 
\beq \al ~=~ \frac{1}{2} ~(1 ~-~ \frac{1}{\sqrt 5}) ~\simeq~ 0.2764, \eeq
where $\mu = \tau$. The filling fraction at this point is $2 \al \simeq 0.5528$.

Finally, we study if there is a $t-V$ model similar to the one discussed in 
Sec.~\ref{sec6} which reduces to Eq.~\eqref{ham5} in the large interaction limit.
We consider a model of the form 
\beq H_7 ~=~ \sum_{j} ~[~ c_{j}^{\dagger}c_{j+1} ~+~ c_{j+1}^\dagger c_j ~+~ 
V_j ~n_{j} ~n_{j+1}], \label{ham7} \eeq
where we again take $V_j$ to vary with $j$ with period four. We carry out an
analysis of the energy costs for the left and right hand sides of Eq.~\eqref{proori5}
similar to the one shown in Table~\ref{Table VII}. We then find that the
energy costs on the two sides of Eq.~\eqref{proori5} are equal
if $V_1 = V_3 = V$ and $V_2 = V_4 = V'$, where $V, ~V'$ are independent parameters.
(Thus the interactions $V_j$ have a period-two structure rather than
period-four). In the limit $V \to \pm \infty$ and $V' \to \pm \infty$, hopping between
sites $j+1$ and $j+2$ will be allowed if and only if $n_j = n_{j+3}$.

Finally, we note that if $V = V'$, i.e., $V_j = V$ for all $j$, the model in
Eq.~\eqref{ham7} becomes the standard $t-V$ model, and it is exactly solvable by 
the Bethe ansatz for a system with PBC~\cite{sutherland}. However, if $V \ne V'$, 
it is not known if the model is exactly solvable. In the limit $V=V' \to 
\infty$, the model is called the folded $XXZ$ model, and this has also been solved 
by the Bethe ansatz~\cite{lenart_2021,pozsgay_2021}.

\section{Discussion} 
\label{sec8}

We begin by summarizing our main results.
We studied a one-dimensional correlated-hopping model of spinless fermions with terms 
involving four consecutive sites having a few global symmetries. This can be mapped to 
an assisted pair-flipping model with terms involving three consecutive sites. We 
found that this model shows strong HSF in a particle
number basis, and time evolution starting from an arbitrary basis state does not always lead to thermalization. In characterizing the HSF in this model, we found it useful to 
define IS, analogous to the constructions used 
earlier~\cite{menon_1997,HariMenon_1995}. The IS provide us with an exponentially large number of conserved quantities which completely characterize the structure of
the HSF. Using the IS, we determined the 
total number of fragments, the number of frozen states, 
and the growth of the size of the largest fragment with the system size. These results were also verified using  transfer matrix methods and explicit enumerations. 

We found that the energy level spacing distribution of the eigenspectrum is approximately Poissonian, but
the Hamiltonian within the largest fragment shows approximately GOE level statistics. Our study of infinite-temperature autocorrelation functions and entanglement 
dynamics also indicated the non-thermal behavior of our model. Further, the finite-size Mazur bound analysis of infinite-temperature autocorrelation functions near one end
and inside the bulk of the system pointed towards a thermal bulk spectrum with a 
non-thermal 
boundary behavior. We also compared our results with another correlated-hopping model involving four consecutive sites, which has been extensively studied in the context of HSF. Finally, we showed how our correlated-hopping model can be realized in
an experimental setting using a variant of the $t-V$ model of spinless fermions in a
particular limit.

In brief, we have considered a model in which the basis states are products of local two-dimensional Hilbert spaces. In this basis, the Hamiltonian has a 
block 
diagonal structure due to the existence of an infinite number of conserved
quantities given by IS. A given block (fragment) may be typical or atypical.
In a typical fragment, the expectation values of local observables for most 
eigenstates at a particular energy tend to the equilibrium values 
within that fragment. We have called this a subspace-restricted weak ETH.

We end by suggesting possible directions for future research. It would 
be useful to determine exactly how fast different fragments grow with system 
size for an arbitrary filling fraction in our model, similar to Eq.~\eqref{nabc} 
which is known for the diffusing dimer model~\cite{menon_1997}. It would 
also be interesting to understand better the large number of zero energy
states in the largest fragment (in particular, to see if some of them
qualify as many-body scars), and to check if such zero-energy states exist in other 
large fragments as well. The transport properties 
vary significantly in different fragments, and it would be useful to understand this better~\cite{bastianello_2022,brighi_2023,wang_2023,barma_slowrelax}. The behaviors of 
bulk and boundary autocorrelation functions for a typical random thermal state in the thermodynamic limit need to be investigated~\cite{hart2023exact}.
Finally, it would be useful to study the effects of disorder~\cite{tomasi_2019,Herviou_2021} and dissipation~\cite{ghosh_2024} in this model. It would also be interesting to see 
if the concept of IS can be generalized to models where HSF occurs in an entangled basis 
rather than in a product state in the particle number basis.

We expect that our results can be experimentally tested in cold-atom 
platforms~\cite{bloch_2008,kohlert2021experimental} where spinless fermionic chains with
spatially periodic potentials and strong interactions can be realized.
Recently, thermalization in some particular fragments of a model with HSF
has been observed in a Rydberg atom system in one dimension~\cite{zhao_2024}. Another observation of HSF has been reported in a superconducting processor in a system exhibiting Stark many-body localization~\cite{wang2024exploring}.

\vspace{0.6cm}
\centerline{\bf Acknowledgments}
\vspace{0.4cm}

S.A. thanks MHRD, India for financial support through a PMRF.
D.S. thanks SERB, India for funding through Project No. JBR/2020/000043.

\appendix

\section{Calculation of number of fragments}
\label{appml}

In this Appendix, we will show how the number of fragments can be calculated using a
transfer matrix method.
As we have discussed earlier, IS cannot contain the substrings of $1111$ and $010$.
This leads us to construct 
the following $8\times8$ transfer matrix $T(C_{i},C_{j})$ in which the rows and columns $C_{i}$ and $C_{j}$ denote configurations of three consecutive sites
labelled as $111$, $110$, $101$, $100$, $011$, $010$, $001$ and $000$. 
We then have 
\bea
T=\left(\begin{matrix} 0 & 1 & 0 & 0 & 0 & 0 & 0 & 0\\0 & 0 & 1 & 1 & 0 & 0 & 0 & 0\\0 & 0 & 0 & 0 & 1 & 0 & 0 & 0\\0 & 0 & 0 & 0 & 0 & 0 & 1 & 1\\1 & 1 & 0 & 0 & 0 & 0 & 0 & 0\\0 & 0 & 0 & 0 & 0 & 0 & 0 & 0\\0 & 0 & 0 & 0 & 1 & 0 & 0 & 0\\0 & 0 & 0 & 0 & 0 & 0 & 1 & 1
\end{matrix}\right). \label{trsec} \eea
The eigenvalues of $T$ are analytically found to be
$\tau$, $-1/\tau$, $\pm i$ and $0$ (which has an algebraic multiplicity of 4), where 
$\tau=(\sqrt{5}+1)/2 \simeq 1.618$ is the golden ratio. (The four non-zero eigenvalues are the roots of the quartic equation $z^{4}-z^{3}-z-1=0$). 

Table \ref{Table I} shows that $M(n)=M(n-1)+M(n-2)$ for $n\geq 4$. We can use this recursion relation to show that 
\bea M(n) ~=~ \tau^{n+1}+\left(-\frac{1}{\tau}\right)^{n+1}, \label{sec} \eea
for $n\geq2$. (Note that only two of the non-zero eigenvalues of $T$ appear in 
Eq.~\eqref{sec}). In fact, Eq.~\eqref{sec} holds even for $n=0$. But for $n=1$, the right hand side of Eq.~\eqref{sec} gives $3$, while a simple counting shows that
$M(1)$ is equal to $2$. 
We now define the generating function for $M(n)$ as
\bea
G(z)=\sum_{n=0}^{\infty}M(n)z^{n}.\label{gen1}
\eea
Using the values of $M(n)$ given above, and summing the series in Eq.~\eqref{gen1}
we obtain
\bea G(z) ~=~ \frac{\tau}{1-z\tau}-\frac{1}{\tau+z}-z. \label{gen3} \eea
Given the values of $M(n)$ in Table \ref{Table I} and Eq.~\eqref{sec}, we 
find that the number of fragments for a system with $L$ sites and OBC is 

\begin{figure}[h!]
\includegraphics[width=\columnwidth]{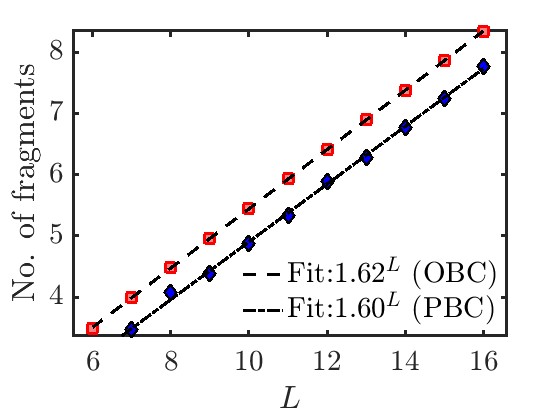}%
\caption{Number of fragments (on a log scale) versus $L$ for OBC and periodic boundary condition (PBC). The numerical fittings show that the numbers 
grow approximately as $1.62^{L}$ and $1.60^{L}$ for OBC and PBC, respectively.} \label{frag} \end{figure}

In Fig. \ref{frag}, we plot the number of fragments $N_{frag}$ versus $L$ 
for both OBC and PBC. In both cases, we see that $N_{frag}$ grows exponentially
as $1.62^{L}$ and $1.60^L$, which are consistent with the analytically estimated 
value of $\tau^L$ for OBC. 

We note that in our numerical work, positive random hoppings 
uniformly distributed in the range $[1,2]$ have been used 
while counting the total number of fragments with PBC in order to avoid any accidental cancellations of sums of matrix elements of the Hamiltonian. As an example of
an accidental cancellation, consider $L=4$ with PBC. Then the state $| 1 0 1 0 \ra$ 
(which denotes the occupation numbers at
sites $1, ~2, ~3, ~4$) can go to $|1 1 1 1 \ra$ in two possible ways, by the
action of either $n_3 d_2^\dagger d_4^\dagger$ or $n_1 d_4^\dagger d_2^\dagger$.
These two terms cancel each other due to the anticommutation relation $d_2^\dagger d_4^\dagger + d_4^\dagger d_2^\dagger = 0$, which would imply that
$\ket{1010}$ and $\ket{1111}$ belong to different fragments.

\section{Calculation of number of frozen states}
\label{appfrozen}

In this Appendix, we will show how the number of frozen states can be calculated. 
Unlike the matrix $T$ defined in 
Eq.~\eqref{trsec} which is designed to remove the configurations $010$
and $1111$, we now need to remove the configurations $010$ and $111$ in order
to find states which are not connected to any other states (and are therefore
frozen). We find that the required transfer matrix is a $4 \times 4$ matrix whose
rows and columns are labelled as $11, ~10, ~01$ and $00$. The required matrix is 
then
\beq T_1 ~=~ \left( \begin{array}{cccc}
0 & 1 & 0 & 0 \\
0 & 0 & 1 & 1 \\
1 & 0 & 0 & 0 \\
0 & 0 & 1 & 1 
\end{array} \right). \label{trmat1} \eeq
We discover that one of the eigenvalues of $T_1$ is zero, while the other 
three are solutions of the cubic equation
\beq \lambda^3 - \lambda^2 - 1 = 0. \label{lambda1} \eeq
The solutions of this equation are given by
\beq \lambda ~=~ \frac{1}{3} ~+~ \frac{2}{3} ~\cos ~\left[ \frac{1}{3} 
\cos^{-1} \left( \frac{29}{2} \right) ~-~ \frac{2 \pi k}{3} \right], 
\label{lambda2} \eeq
where $k$ can take the values $0, ~1, ~2$. We then find the three 
eigenvalues to be $1.466$ and $- 0.233 \pm 0.793 i$ approximately; the
magnitudes of the last two eigenvalues are less than 1.
Hence, the number of frozen states increases asymptotically as $1.466^L$. 
Note that this is a slower growth than the total 
number of fragments which increases as $1.618^L$.


The number of frozen states can also be counted using the fact that the number of such states, $N_{frozen}(L)$, with OBC is given by the sum of all the 
matrix elements of $T_{1}^{L-2}$, for $L \geq 3$. Defining $Y=T_{1}^{L-2}$, we have
\bea N_{frozen} ~=~ \sum_{i,j=1}^{4} Y_{i,j}. \eea

The counting of frozen states with PBC is slightly different since we have
to take care of the constraint
that the states of the four consecutive sites ($L-1,L,
1,2)$ should not contain either $111$ or $010$. Therefore, defining $Y=T_{1}^{L-2}$ and taking into account this additional constraint, we find that
\bea N_{frozen}(L) &=& Y (1, 2) + Y (1, 4) + Y (2, 3)\non\\&& + Y(3,1)+ Y (3,2)+ Y(3,4)\non\\&&+Y (4, 1) +Y(4,3)+ Y (4, 4). \label{Nper} \eea

The number of frozen states for first few system sizes shown in 
Table \ref{Table III} exactly agree with the numerically 
obtained numbers. We note again that for PBC, we have used random hoppings which 
are uniformly distributed in the range $[1,2]$ to numerically compute the 
number of frozen states to avoid any accidental cancellations between 
different matrix elements of the Hamiltonian.

\section{Calculation of the size of the largest fragment of the three-spin model}
\label{appA}

We have seen that there are exponentially many frozen fragments which have only one state each. Fragments containing $n$ blinkers have size $2^n$. However, there 
are other fragments which are much larger in size. We typically find that larger
fragments correspond to IS with shorter length (see Fig.~\ref{ISsize}). We will 
now study the largest fragments whose IS turn out to consist of either the null 
string ($\phi$), 1, $11$ or $111$.

We will use the method of enumerative combinatorics of characters~\cite{generating1994,HariMenon_1995, menon_1997, Barma_1994} to evaluate 
the size of the fragments of the three-spin model with OBC whose IS consist of 
only $\phi$ or 1's. For clarity, we will use the symbols $A$ and $B$, rather than 
$0$ and $1$, to denote the two characters. 

We define four formal infinite sums, $U_{0}$, $U_{1}$, $U_{2}$ and 
$U_{3}$ as sums of all distinct strings made of characters $A$ and $B$ that correspond to IS which can be $\phi$, $B$, $B^2$ or $B^{3}$, respectively. We assign a weight $x$ and $y$ to each occurrence of $A$ and $B$, and the weight of a string having $p$ number of $B$'s and $q$ number of $A$'s is 
$x^{p}\,y^{q}$. We denote the sum of weights of these formal series by $U_{0}(x,y)$, $U_{1}(x,y)$, $U_{2}(x,y)$ and $U_{3}(x,y)$. 

In the sum of terms $U_{0}(x,y)$, there are terms is which contain no $A$'s
and only $4r$ $B$'s, and other terms, which have an even number of $A$'s, must have the following structure:
\bea U_{0} ~=~ \frac{\phi}{1-B^4} ~+ \sum_{r=0,1,2,3} \sum_X ~\frac{B^{r}}{1-B^4} ~AWA~ X,\label{nullstring} \eea
where $W$ is the sum over all possible substrings of between $A$'s that reduce to $B$, so that $AWA$ reduces to $ABA$ and therefore to $B^3$. Moreover, $X$ must be reducible to $B^{s}$ such that $r+3+s \equiv 0$ mod 4.
Then writing the different possibilities of $r$ mod 4 explicitly, we obtain
\bea U_{0}&=&\frac{\phi}{1-B^4} ~+~ \frac{\phi}{1-B^4} ~AWA~ U_{1} \non \\
&&+ ~\frac{B}{1-B^4} ~AWA~ U_{0} ~+~ \frac{B^2}{1-B^4} ~AWA ~ U_{3} \non \\
&&+ ~\frac{B^3}{1-B^4} ~AWA ~U_{2}. \label{U0} \eea

Then the generating function $U_0$ in Eq.~\eqref{U0} is given by
\bea U_{0}(x,y)&=& \frac{1}{1-x^4} ~+~ \frac{y^2}{1-x^4} ~W(x,y)\,U_{1}(x,y) 
\non \\
&&+ ~\frac{xy^2}{1-x^4} ~W(x,y)\,U_{0}(x,y) \non \\
&& +~ \frac{x^2 y^2}{1-x^4} ~W(x,y)\,U_{3}(x,y) \non \\
&& +~ \frac{x^{3}y^{2}}{1-x^4} ~W(x,y)\,U_{2}(x,y), \label{genU0} \eea
where $W(x,y)$ denotes the weight of $W$. Since $AWA$ must reduce to $B^3$, 
it must contain an even number of $A$'s, and $y^2 W(x,y)$
must have terms of the form $x^r y^{2s}$ where $r+2s=3$ mod 4, $r \ge 1$,
and $s \ge 1$. Hence we can write 
\beq y^2 W(x,y) = x y^2 H(x,y), \label{hxy} \eeq
where $H(x,y)$ has terms of the form $x^p y^{2q}$ where $p+2q = 0$ mod 4.

In a similar manner, one can show that the generating functions
$U_{1}$, $U_{2}$ and $U_{3}$ can be written as
\bea U_{1}(x,y)&=&\frac{x}{1-x^4} ~+~ \frac{xy^2}{1-x^4} ~W(x,y)\,U_{1}(x,y)
\non\\
&& + ~\frac{y^2}{1-x^4}~ W(x,y)\,U_{2}(x,y) \non \\
&& +~ \frac{x^{3}y^2}{1-x^4} ~W(x,y)\,U_{3}(x,y) \non \\
&& +~ \frac{x^{2}y^{2}}{1-x^4} ~W(x,y)\,U_{0}(x,y), \label{genU1} \eea
\bea
U_{2}(x,y)&=&\frac{x^2}{1-x^4} ~+~ \frac{x^{3}y^2}{1-x^4}~ W(x,y)\,U_{0}(x,y)
\non\\
&& +~ \frac{x^{2}y^2}{1-x^4} ~W(x,y)U_{1}(x,y) \non \\
&& +~ \frac{x y^2}{1-x^4} ~W(x,y)\,U_{2}(x,y) \non \\
&& +~ \frac{y^{2}}{1-x^4} ~W(x,y)\,U_{3}(x,y), \label{genU2} \eea
\bea
U_{3}(x,y)&=&\frac{x^3}{1-x^4} ~+~ \frac{y^2}{1-x^4} ~W(x,y)\,U_{0}(x,y)\non\\
&& + ~\frac{x^{3}y^2}{1-x^4} ~W(x,y)\,U_{1}(x,y) \non \\
&& +~ \frac{x^{2} y^2}{1-x^4} ~W(x,y)\,U_{2}(x,y) \non \\
&& +~ \frac{xy^{2}}{1-x^4} ~W(x,y)\,U_{3}(x,y). \label{genU3} \eea

The generating functions given above can be combined as 
\beq V(x,y) = U_{0}(x,y) + U_{1}(x,y) + U_{2}(x,y) + U_{3}(x,y). \eeq
Using the identity $(1 + x + x^2 + x^3)/(1 - x^4) = 1/(1-x)$ and
Eq.~\eqref{hxy}, we find that
\beq V(x,y) ~=~ \frac{1}{1 ~-~ x ~-~ xy^2 ~H(x,y)}. \label{vhxy} \eeq
We now set $y=x$. Then $U_0, ~U_1, ~U_2, ~U_3$ will become polynomials in
$x$ with terms whose degrees are equal to $0, ~1, ~2, ~3$ mod 4, respectively,
and $H(x,y) = H(x)$ will become a polynomial in $x^4$. We then obtain
\bea V(x)&=&\frac{1}{1-x-x^3\,H(x)},\label{gensl} \eea
where $V(x)$ is the generating function of all strings that reduce to 
$B$'s only, and $xH(x)$ is related to IS which reduce to a single $B$.
By direct examination of strings of lengths $0-3$ (namely, $\phi \to 1$,
$B \to x$, $BB \to x^2$, $BBB \to x^3$ and $ABA \to xy^2 = x^3$ for $y=x$),
we find that the first few terms in $V(x)$ and $H(x)$ are given by
\bea V(x) &=& 1 ~+~ x ~+~ x^2 ~+~ 2x^3 ~+~ \cdots, \non \\
H(x) &=& 1 ~+~ \cdots. \label{vhx} \eea

Next, we can write $V(x)$ as
\bea V(x)=V_{0}(x)+x\,V_{1}(x)+x^2\,V_{2}(x)+\,x^{3}\,V_{3}(x),\label{V} \eea
where $V_{0}$, $V_{1}$, $V_{2}$ and $V_{3}$ are all polynomials in $x^{4}$.
Using Eq.~\eqref{gensl} and the fact that $V_{0}$, $V_{1}$, $V_{2}$ and $V_{3}$ are polynomials in $x^{4}$, we can show that
\bea V_{0}(x) &=& \frac{V(x) ~+~ V(-x) ~+~ V(ix) ~+~ V(-ix)}{4}, \non \\
&=&\frac{1-2x^{4}H}{(1-2x^4 H)^2-x^4(1+x^{4}H^2)^2},\label{V0}\\
V_{1}(x)&=& \frac{V(x) ~-~ V(-x) ~-~ i V(ix) ~+~ i V(-ix)}{4x}, \non \\
&=&\frac{1-x^4+x^8\,H^3}{(1-2x^4 H)^2-x^4(1+x^{4}H^2)^2}, \label{V1} \non \\
V_{2}(x) &=& \frac{V(x) ~+~ V(-x) ~-~ V(ix) ~-~ V(-ix)}{4x^2}, \non \\
&=&\frac{1+x^4 H^2}{(1-2x^4 H)^2-x^4(1+x^{4}H^2)^2}, \label{V2} \eea
\bea V_{3}(x)&=& \frac{V(x) ~-~ V(-x) ~+~ i V(ix) ~-~ i V(-ix)}{4x^3}, \non \\
&=&\frac{1+H-x^4\,H^2}{(1-2x^4 H)^2-x^4(1+x^{4}H^2)^2} \label{V3}. \eea

Next, we can show that $V_{3} (x) =2V_{2} (x)$ as follows.

\noi 1. Given a string in $V_{2}$,
one can add a $1$ to the left to obtain a string in $V_{3}$.
Further, this process is reversible, 
i.e., given a string in $V_{3}$ beginning with $1$, one can delete
the 1 to obtain a string in $V_{2}$.

\noi 2. Next, given a string belonging to $V_{2}$ 
that begins with a $0$, one can add a 1 on the right side of the string to 
obtain a string in $V_{3}$.
This process is also reversible: given a string in $V_3$ which begins
with a 0 and ends with a 1, we can delete the 1 to obtain a string in $V_2$
which begins with a 0.

\noi 3. Finally, given a string in $V_{2}$, 
that begins with a 1, we can replace the 1 by a 0 and add a 0 at the right 
end of the string. This procedure thus produces a string in $V_{3}$.
This mapping is again reversible similar to rules 1 and 2.

Taking $V_{3}=2V_{2}$ into account and using Eq.~\eqref{V2} and \eqref{V3}, we 
obtain the following quadratic equation for $H(x)$
\bea 3x^4\,H^2-H+1 &=& 0\,,\non\\
{\rm which ~implies} ~~~ H(x) &=& \frac{1 ~\pm~ \sqrt{1-12x^4}}{6x^4}. 
\label{H} \eea
Eq.~\eqref{vhx} tells us that we have to take the lower sign (minus) in the
expression for $H(x)$ in Eq.~\eqref{H}. We then obtain the final expression 
\bea V(x)=\frac{1}{1 ~-~ x ~-~ \left(\frac{1 ~-~ \sqrt{1-12x^4}}{6x}\right)}. \eea
This is Eq.~\eqref{Vexp} in the main text.

\section{Comparison between the growths of the largest fragments in four-site and three-site models}
\label{appB}

In this section, we compare the sizes of the fragments generated from the root state, $1010 \cdots 10$ (or $0101 \cdots 01$) of the four-site model with the one 
generated from the root state, $111 \cdots 111$ of the three-site model. Note that both the states $1010 \cdots 10$ and $0101 \cdots 01$ of the four-site model 
map to the same state, $111 \cdots 111$, of the three-site model under bond-site mapping. We first we compare the cases with OBC. As evinced in Table 
\ref{Table VIII}, the sizes of fragments originating from either of the 
root states $1010 \cdots 10$ and $0101 \cdots 01$ of the four-site model with 
$L$ sites are exactly the same as the one obtained 
from the root state $111 \cdots 111$ of the three-site model with $L-1$ sites. 
This can be anticipated from the fact that the bond-site mapping mentioned in 
Sec.~\ref{sec2}
maps two states of the four-site model with $L$ sites to a single state of the 
three-site model with $L-1$ sites with OBC. 

\begin{table}[htb]
\begin{center}
\begin{tabular}{|c|c|c|} 
\hline
$L$& $~D^{\text{four-site}}_L~$ & $~D^{\text{three-site}}_L~$ \\
\hline
3&1&2\\
\hline
4&2&3 \\
\hline
5&3&4\\
\hline
6&4&6\\
\hline
7&12&12\\
\hline
8 &12&19\\
\hline
9&19&28\\
\hline
10&28&46\\
\hline
11&46&92\\
\hline
12&92&150\\
\hline
13&150&232\\
\hline
14&232&396\\
\hline
15&396&792\\
\hline
16&792&1315\\
\hline
17&1315&2092\\
\hline
18&2092&3646\\
\hline
19&3646&7292\\
\hline
20&7292&12258\\
\hline
21&12258&19864\\
\hline
22&19864&35076\\
\hline
23&35076&70152\\
\hline
24&70152&118990\\
\hline
\end{tabular}
\caption{Sizes of fragments generated from the
root states $1010 \cdots 10$ (or $0101 \cdots 01$) and $111 \cdots 11$ for the 
four-site and three-site models respectively, with OBC. We observe that two 
fragments of identical sizes of the four-site model with $L$ sites maps to a single 
fragment originated from the root state $1111 \cdots 11$ of the three-site model
with $L-1$ sites.} \label{Table VIII}
\end{center}
\end{table}

Next, we compare the fragment sizes for the four-site and three-site models with 
PBC. We will assume that the system size $L$ is even to take the periodicity of
the four-site model into account. As shown in Table 
\ref{Table IX}, the size of the fragment for the root state
$1010 \cdots 10$ (or $0101 \cdots 01)$ of the four-site model is two times larger than that of the three-site model for the root state $111 \cdots 11$ for $L=4n$. 
This is due to the fact that the global symmetries $C_{2}$ and $C_{3}$ of the 
four-site model, as 
discussed in Eq.~\eqref{globalsym}, are well-defined for $L=4n$, and hence, the states $1010 \cdots 10$ and $0101 \cdots 01$ lie in the same symmetry sector and 
belong to a single fragment for a system with PBC. On the other hand, we find that
two different fragments arise from the root states $1010 \cdots 10$ and $0101 \cdots 
01$ in the four-site model whose sizes are the same as the single fragment 
of the three-site model for a chain with $L=4n+2$ sites with PBC. We also note
that the two root states of the four-site model are not connected to each other by the global symmetries $C_{2}$ and $C_{3}$ for $L=4n+2$, and therefore they 
generate two different fragments of identical sizes.

\begin{table}[htb]
\begin{center}
\begin{tabular}{|c|c|c|} 
\hline
 $L$& $~D^{\text{four-site}}_L~$ & $~D^{\text{three-site}}_L~$\\
\hline
4&6&3 \\
\hline
6&10 &10\\
\hline
8 &38&19\\
\hline
10&106&106\\
\hline
12&300&150\\
\hline
14&1156&1156\\
\hline
16&2630&1315\\
\hline
18&12826&12826\\
\hline
20&24516&12258\\
\hline
22&143980&143980\\
\hline
24&237980&118990\\
\hline
26&1630084&1630084\\
\hline
\end{tabular}
\caption{sizes of fragments generated from the root configurations $1010 \cdots$
(or $0101 \cdots 01$) and $111 \cdots 111$ for the four-site and three-site model,
respectively, with PBC. We note that for $L=4n$, the two states belong to the same 
fragment in the four-site model, whereas for $L=4n+2$, the four-site 
model exhibits two different fragments with the same size $D_L$, one coming from the root configuration $101010 \cdots 10$ and the other from $010101 \cdots 01$.}
\label{Table IX}
\end{center}
\end{table}

\section{Fragmentation structure of the four-site model away from half-filling}
\label{appC}

We will discuss here strong HSF away from half-filling in our model. In doing so,
we consider the following root states of the four-site model with OBC, 
$\cdots 011110011110 \cdots$ or $\cdots 10001100001 \cdots$ which correspond to
filling fractions $\nu=2/3$ and $1/3$ respectively. Both these states map to the state $\cdots 101000101000 \cdots$ in the three-site model with 
OBC. Taking this state as a root configuration of the three-site model, we 
find the dimension of this Hilbert space fragment by numerical 
enumeration for $L=6, ~12, ~18$ and $24$; the results are shown in
Table ~\ref{Table X}. We have chosen the system sizes to be 
multiples of 6 to ensure that the pattern of the root state remains invariant
as the system size is increased.

\begin{center}
\begin{table}[htb]
\begin{tabular}{|c|c|c|c|c|c|c|c|c|c|c|c|c|c|c|c|} 
\hline
$L$&6&12&18&24\\
\hline
$D_{i,L}$&7&107&1906&35259\\
\hline
\end{tabular}
\caption{Size of the fragment $D_{i,L}$ versus $L$ obtained from the 
root state $101000101000 \cdots$ for the three-site model with OBC.}
\label{Table X}
\end{table}
\end{center}

\begin{figure}[htb]
\includegraphics[width=\columnwidth]{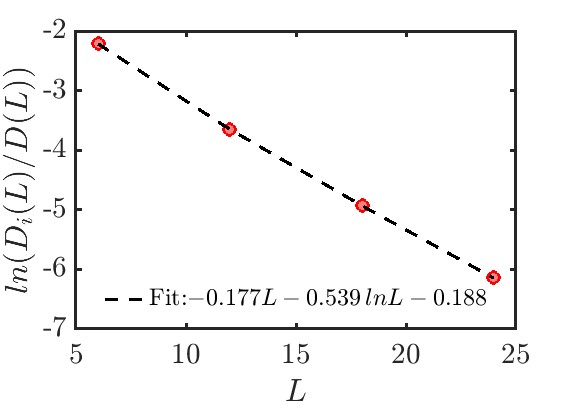}%
\caption{Plot of the ratio of the size $D_i$ of the fragment of interest 
to the total dimension $D = 2^L$ of the Hilbert space (shown on a log scale) 
versus $L$ for a system with with OBC. The ratio is seen to decrease 
exponentially with $L$.} \label{scalingawayhalf} \end{figure}

\begin{figure}[htb]
\stackon{\includegraphics[width=0.514\linewidth]{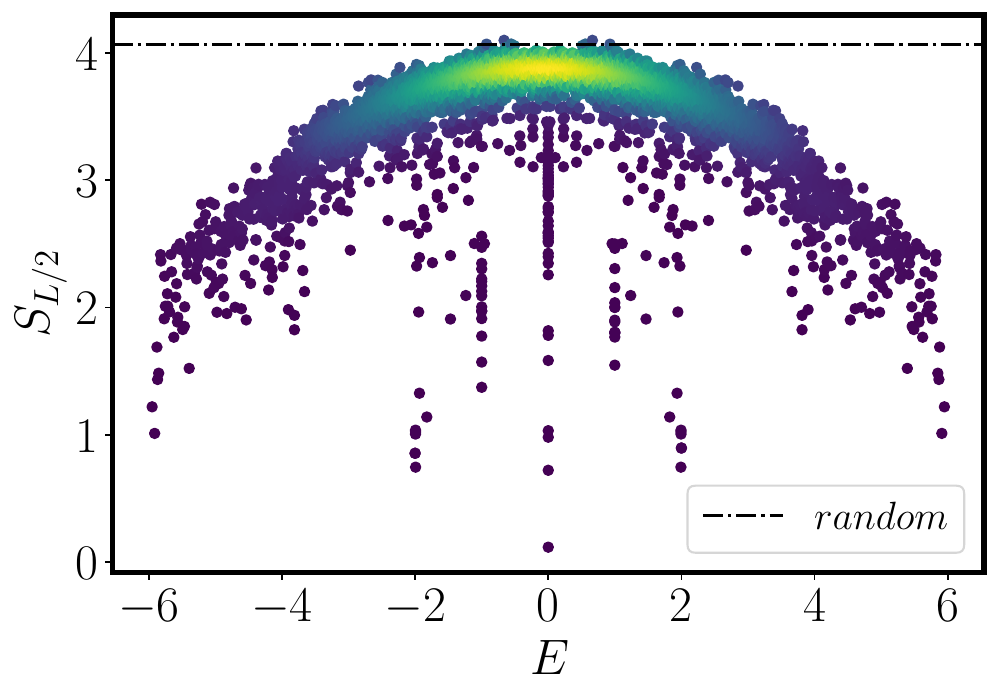}}{(a)}%
\stackon{\includegraphics[width=0.514\linewidth]{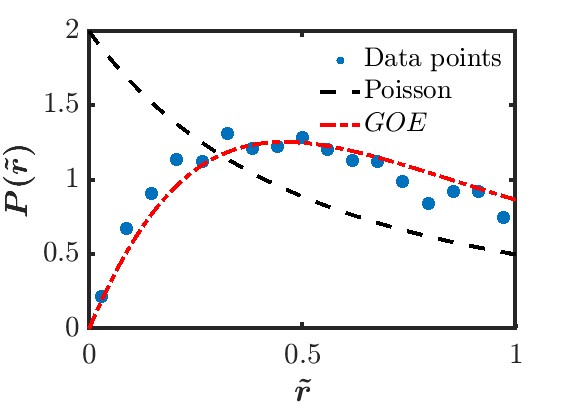}}{(b)}
\caption{(a) Plot of $S_{L/2}$ versus $E$ for the fragment generated from the root state, $0111001110 \cdots$ for $L=24$ at 
filling fraction $\nu=2/3$ for the four-site model with OBC. The fragment
size is $4906$. The dash-dot line indicates the value of the entanglement 
entropy of a random state in this subspace. The entanglement spectrum contains a small fraction of eigenstates with low entanglement entropy in the middle of the 
spectrum. (b) The level spacing ratio analysis for the same fragment in the presence of a small uniformly distributed random disorder with disorder
strength $w=0.01$. The probability distribution of $\Tilde{r}$ is close to the GOE distribution with $\langle\Tilde{r}\rangle \simeq 0.51$.}
\label{spectrumaway} \end{figure}

In Fig. \ref{scalingawayhalf}, we show how this fragment grows with $L$ by the numerical enumeration method. Moreover, the numerical fitting indicates 
that the size of this particular fragment grows as $1.68^{L}/L^{0.54}$. This again 
implies that $D_{i}/D$ goes to zero for $L\rightarrow \infty$, where $D$ being 
the total dimension of the Hilbert space, which indicates strong 
HSF~\cite{sala_ergo_2020}. This observation leads us to conclude that our 
four-site model exhibits strong HSF~\cite{sala_ergo_2020} at arbitrary filling
fractions.

In Figs. \ref{spectrumaway} (a-b), we show the half-chain entanglement entropy as a function of the energy $E$ for the fragment generated from the root state $011110011110 \cdots$ at $\nu=2/3$ for the four-site model with $L=24$ and OBC;
the size of this fragment is $4906$. The spectrum shows that most of the eigenstates lie close to a single curve as in a thermal system. Nevertheless, there is also a small fraction of states in the middle of the spectrum with 
anomalously low entanglement entropy. We further perform an analysis of the 
energy level statistics within this fragment after adding a small amount of randomly distributed on-site disorder of strength $w=0.01$ for the same reasons mentioned in Sec.~\ref{sec4}. The probability distribution of $\tilde{r}$ 
within this fragment is found to follow the 
GOE with $\langle\Tilde{r}\rangle \simeq 0.51$, which is close to the GOE value. The fluctuations in the analysis of $\Tilde{r}$ arise due to the 
limited system sizes. Since we want to keep the $01110$ pattern of 
the root state intact while generating this sector, we must take 
$L$ to be a multiple of six. We have to limit our analysis to $L=24$
since the next system size $L=30$ is numerically very difficult.
We find that our model always exhibits strong HSF irrespective of the 
filling unlike models where a transition from strong to weak HSF can occur
as a function of the filling fraction~\cite{Morningstar_2020,wang_2023}.

\newpage

\bibliography{ref}

\end{document}